\documentclass[twocolumn]{autart}    

\usepackage{graphicx}          

\usepackage{comment}

\usepackage{amsmath}
\usepackage{nomencl}
\usepackage{mathtools}
\usepackage{amssymb}
\makenomenclature

\usepackage{url}
\usepackage{xspace}
\usepackage{xfrac}
\usepackage{array}
\usepackage[noend]{algpseudocode}
\usepackage{algorithm}
\renewcommand*\Call[2]{\textproc{#1}(#2)}

\usepackage{siunitx}
\usepackage{subcaption}
\usepackage{pgfplots}
\usepackage{layouts}

\graphicspath{{./img/}}

\newtheorem{defi}{Definition}
\newcommand{\fakepar}[1]{\vspace{1mm}\noindent\textbf{#1.}}

\newcommand{\norm}[1]{\left\lVert#1\right\rVert}

\newcommand{\transp}{\text{T}}

\let\originalleft\left
\let\originalright\right
\renewcommand{\left}{\mathopen{}\mathclose\bgroup\originalleft}
\renewcommand{\right}{\aftergroup\egroup\originalright}

\newcommand\figref[1]{Fig.~\ref{#1}}

\newcommand\secref[1]{Sec.~\ref{#1}}
\newcommand\appref[1]{App.~\ref{#1}}
\renewcommand\algref[1]{Alg.~\ref{#1}}
\renewcommand\eqref[1]{(\ref{#1})}
\newcommand\defref[1]{Def.~\ref{#1}}

\renewcommand{\etal}{et~al.\xspace}
\newcommand{\eg}{e.g.,\xspace}
\newcommand{\ie}{i.e.,\xspace}

\newcommand{\cf}{cf.\xspace}

\newlength{\collen}
\usepackage[toc,section=chapter,nonumberlist]{glossaries}
\glsdisablehyper

\usepackage{ifthen}
\newboolean{authnotes}

\setboolean{authnotes}{true}

\ifthenelse{\boolean{authnotes}}
{
\newcommand{\todo}[1]{{{\bf\color{magenta} TODO: #1}}}
\newcommand{\nf}[1]{\footnote{{\bf\color{blue} Niklas: #1}}}
\newcommand{\db}[1]{\footnote{{\bf\color{green!50!black} Dominik: #1}}}
\newcommand{\st}[1]{\footnote{{\bf\color{purple!90!black} Sebastian: #1}}}
}
{
\newcommand{\todo}[1]{}
\newcommand{\nf}[1]{}
\newcommand{\db}[1]{}
\newcommand{\st}[1]{}
}

\newacronym{etc}{ETC}{event-triggered control}
\newacronym{et}{event-triggered}{event-triggered}
\newacronym{rl}{RL}{reinforcement learning}
\newacronym{nn}{NN}{neural network}
\newacronym{zoh}{ZOH}{zero-order hold}
\newacronym{relu}{ReLU}{rectified linear unit}
\newacronym{ppoc}{PPOC}{proximal policy option-critic}
\newacronym{ppo}{PPO}{proximal policy optimization}
\newacronym{trpo}{TRPO}{trust region policy optimization}
\newacronym{gae}{GAE}{generalized advantage estimation}
\newacronym{tanh}{TanH}{hyperbolic tangent}
\newacronym{dof}{DoF}{degrees of freedom}
\glsunset{et}

\begin{document}

\begin{frontmatter}
\title{Learning Event-triggered Control from Data \\ through Joint Optimization}

\author[MPIa]{Niklas Funk}\ead{nwfunk@gmx.net},    
\author[RWTH,MPIa]{Dominik Baumann}\ead{dbaumann@tuebingen.mpg.de},               
\author[MPIb]{Vincent Berenz}\ead{vberenz@tuebingen.mpg.de},               
\author[RWTH,MPIa]{Sebastian Trimpe}\ead{trimpe@dsme.rwth-aachen.de}  

\address[MPIa]{Intelligent Control Systems Group, Max Planck Institute for Intelligent Systems, Stuttgart, Germany}
\address[MPIb]{Empirical Inference Department, Max Planck Institute for Intelligent Systems, T\"ubingen, Germany}                                           
\address[RWTH]{Institute for Data Science in Mechanical Engineering, RWTH Aachen University, Aachen, Germany}                                             
          
\begin{keyword}                           
Event-triggered Control; Reinforcement Learning; Stability Verification; Neural Networks              
\end{keyword}                             

\begin{abstract}
We present a framework for model-free learning of event-triggered control strategies. 
Event-triggered methods aim to achieve high control performance while only closing the feedback loop when needed.
This enables resource savings, \eg network bandwidth if control commands are sent via communication networks, as in networked control systems.
Event-triggered controllers consist of a communication policy, determining when to communicate, and a control policy, deciding what to communicate.  
It is essential to jointly optimize the two policies since individual optimization does not necessarily yield the overall optimal solution.
To address this need for joint optimization, we propose a novel algorithm based on hierarchical reinforcement learning.
The resulting algorithm is shown to accomplish high-performance control in line with resource savings and scales seamlessly to nonlinear and high-dimensional systems. 
The method's applicability to real-world scenarios is demonstrated through experiments on a six degrees of freedom real-time controlled manipulator. Further, we propose an approach towards evaluating the stability of the learned neural network policies.
\end{abstract}

\end{frontmatter}

\footnotetext[1]{\textit{DOI:} \url{https://doi.org/10.1016/j.ifacsc.2021.100144}}


\section{Introduction}
\label{chp:Introduction}

In modern control systems, control commands often need to be transmitted over (wired or wireless) communication networks~\cite{hespanha2007survey,lunze2014control}.
Examples of such networked control systems include swarms of drones, where communication is needed for drones to fly in formation; autonomous cars, where exchanging information between vehicles may increase traffic throughput and reduce fuel consumption; or smart homes, where distributed sensors, actuators, and computing units need to cooperate to regulate the indoor climate.
In all examples, multiple systems utilize the same network for communication.
If all systems transmit their information at high periodic rates, this can easily overload the network and result in an increased probability of message loss and longer transmission delays~\cite{7349192}.
Further, in many applications, distributed sensors and computing units should be untethered and thus battery-driven. In these cases, limiting communication can significantly improve battery life.
\Gls{et} methods have been developed explicitly to serve this need of controlling systems at reduced communication rates, see for instance~\cite{heemels2012introduction,miskowicz2018event,grune2014event,lemmon2010event} for an overview.

\begin{figure}
\centering
\includegraphics[width=1.0\collen]{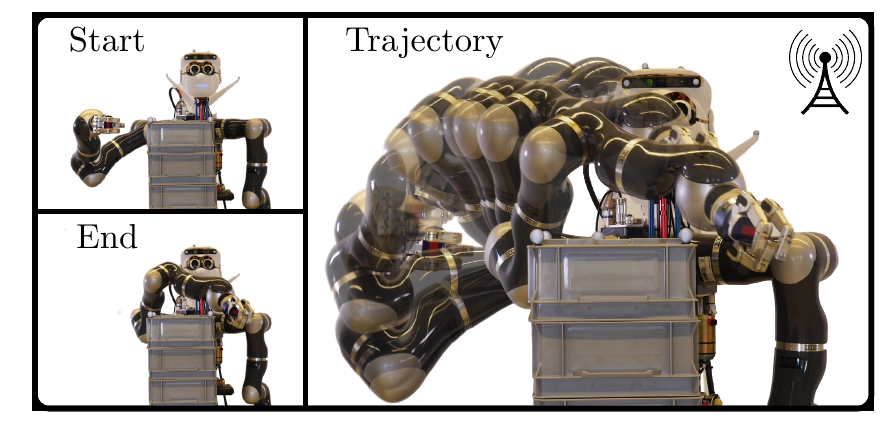}
\caption{Performing event-triggered control on the Apollo robot. The overlayed frames coincide with the time instants in which a new control command is computed and applied. The robot successfully avoids the obstacle with only a few recomputations of the control signal and thus saving \SI{90}{\percent} of communication.}
\label{fig:title_hw_apollo}
\end{figure}

In \gls{etc}, closing the feedback loop and, thus, transmitting information, is triggered by the occurrence of certain events, \eg an error growing too large.
Practical investigations have shown that \gls{etc} can significantly reduce the amount of communication while still achieving high-performance control~\cite{trimpe2011experimental,araujo2013system,dolk2017event}.
In most \gls{et} approaches, the design of the control and communication strategy is based on a known mathematical model of the system~\cite{heemels2012introduction}.
Yet, for complex systems, an accurate description may not be readily available.
Further, the majority of results only consider linear and low-dimensional systems.
Available results for nonlinear systems mostly optimize the  communication and control policies separately or fix one of both~\cite{demirel2018deepcas,vamvoudakis2018model}.
This is problematic as in \gls{etc}, the separation principle does not hold in general~\cite{ramesh2011dual}. That is: even if both policies are individually optimal, their combination does not necessarily yield the overall optimal solution.
For optimal \gls{etc}, the control and communication policy need to be optimized \emph{jointly}.
In this article, we propose an algorithm based on model-free \gls{rl} that jointly learns the control and communication policies from data. By exploiting model-free algorithms, we mitigate the need for an accurate dynamics model, and since we do not make any assumptions on dimensionality or linearity, the resulting framework can equally be applied to linear and nonlinear, low- and high-dimensional systems.

A key hurdle of learning \gls{etc} with \gls{rl} is the hybrid action space of such controllers.
At each time step, the controller takes a discrete decision, whether or not to communicate. In case of communication, a continuous control input is transmitted.
Hierarchical \gls{rl}~\cite{PrecupPHD} naturally captures this hybrid decision structure.
It provides a top-level policy that decides which action to take (in our case, whether or not to communicate). Depending on this choice, the corresponding low-level policy is evaluated, which, for instance, yields the control input in case of communication. To the best of our knowledge, hierarchical \gls{rl} has not yet been used for learning \gls{etc}. 
However, to successfully learn \gls{et} controllers, it is not sufficient to apply the existing hierarchical algorithms. This is mainly because the \gls{etc} setting restricts exploration. Unlike in periodic control, in \gls{etc}, varying the control action for exploration is only possible on communication
instances. Thus, exploration directly conflicts with the goal of saving communication. We extend the hierarchical algorithms accordingly, which results in the first method that successfully derives \gls{etc} strategies through joint optimization in nonlinear and high-dimensional environments.


One of the drawbacks of learning-based approaches, especially those based on \glspl{nn}, is that they often do not provide stability guarantees. However, such statements are crucial when deploying the controllers at scale in potentially safety-critical, real-world applications.
We approach this challenge by proposing a verification framework, capable of checking the stability of the learned policy and refining it if necessary. 
The framework combines output range analysis of the learned control policy with model knowledge.
That way, we can utilize a popular \gls{nn} verification framework~\cite{katz2017reluplex} to provide stability guarantees based on control invariant sets for linear systems.

\fakepar{Contributions} 
We make the following contributions:
\begin{itemize}
\item Leveraging and extending hierarchical \gls{rl} algorithms to obtain resource-aware, \gls{et} controllers; 
\item Presenting an algorithm, capable of end-to-end learning of the control and communication policy through joint optimization for high-dimensional nonlinear systems;
\item Demonstrating the algorithm's practical significance by applying learned \gls{etc} strategies on a real robotic system, as illustrated in \figref{fig:title_hw_apollo};
\item Presenting an algorithm to check the stability of linear \gls{et} systems controlled by \gls{nn} policies with \gls{relu} activations. We also provide a method for refining the \glspl{nn} in case of initial instability.
\end{itemize}

\fakepar{Outline}
We start with an overview of related work before we provide the problem formulation and necessary background.
Next, we introduce the developed algorithm and present results in challenging, high-dimensional simulation environments and on a real robotic system.
Lastly, we discuss the stability verification procedure and conclude with a discussion.

\section{Related Work}
\label{chp:RelatedWork}


\fakepar{\gls{etc} overview} For a general introduction and overview of \gls{etc}, we refer the reader to~\cite{lunze2014control,heemels2012introduction,miskowicz2018event,lemmon2010event}. 
While~\cite{heemels2012introduction} focuses on linear systems, the other references also discuss approaches for event-triggered control of nonlinear systems.
All methods, regardless of whether they are designed for linear or nonlinear systems, rely on an accurate model of the system's dynamics.
In practical applications, especially when considering complex nonlinear systems, such a model may not be available.
We address the more general and more challenging problem of learning event-triggered control and communication policies without assuming any knowledge of the system dynamics. 
\fakepar{Learning \gls{etc} -- imposing structure} Recently, there have been several other works that learn \gls{et} communication and control policies in a model-free way. In \cite{demirel2018deepcas}, the authors propose to use \gls{rl} to learn a scheduling strategy for controlling a multi-agent system. 
The approach focuses on arbitrating the constrained communication bandwidth among the agents, while the agents' control strategy is fixed beforehand and not subject to active optimization. 
Vamvoudakis et al. \cite{vamvoudakis2018model} exploit concepts from Q-learning, and propose a model-free algorithm that comes up with an \gls{etc} strategy for linear systems in continuous time. 
In their approach, the general triggering condition is predefined. Therefore, the optimization algorithm only operates on the triggering threshold of the communication strategy (\ie when to trigger). 
The authors of \cite{zhong2014event} and \cite{yang2017event} also use the same predefined triggering condition, but derive \gls{etc} strategies for nonlinear systems. 
While \cite{zhong2014event} uses an adaptive dynamic programming approach, \cite{yang2017event} relies on an identifier-critic architecture, actively identifying the unknown system dynamics. 
The authors of \cite{sahoo2015neural} use \glspl{nn} to parametrize the control policy, while the triggering condition depends on the state and the weights of the controller. 
Further, the \gls{nn} is updated in an aperiodic fashion, \ie only on triggering instances. 
The method proposed herein is more general. We jointly learn the control and communication strategy from scratch without imposing any additional structure, \eg concerning the triggering condition. This is especially crucial as the separation principle does not hold in general for \gls{etc} \cite{ramesh2011dual}. Moreover, we showcase the performance of our algorithm in substantially more complex and higher-dimensional simulation environments than those presented in the above references, as well as on a real 6 \gls{dof} real-time controlled manipulator.

\fakepar{Learning \gls{etc} -- without imposing structure} There are only few other approaches~\cite{BaumannRL,hashimoto2019learning} that learn ETC without a-priori restrictions on the triggering condition. Baumann et al.~\cite{BaumannRL} discuss both a joint optimization of control and communication and a separate optimization procedure.
However, only the separate optimization can be applied to high-dimensional tasks.
The authors of~\cite{hashimoto2019learning} propose to use Gaussian processes to learn a model of an unknown system, which is then exploited to derive an optimal, self-triggered control strategy through approximated value iteration.
Due to the computational complexity, this approach is limited to low-dimensional systems, and a maximum inter-communication time needs to be fixed a priori.
Thus, neither of those approaches can jointly optimize the control and communication policy in high-dimensional settings, as we do herein.
Further, neither work provides any stability guarantees or results on real hardware.

\fakepar{Deep \gls{rl} for learning control} Learning control policies from data for high-dimensional and nonlinear systems has been studied extensively in recent years~\cite{lillicrap2015continuous,duan2016benchmarking,levine2016end,peters2008reinforcement,kober2013reinforcement}.
However, all these works consider periodic communication and are, therefore, not applicable to the problem considered herein.
\gls{etc} is a challenging problem in that it leads to a hybrid action space.
Approaches that deal with such hybrid action spaces will be discussed next. 
In~\cite{masson2016reinforcement,hausknecht2016deep}, the authors propose to rephrase the problem using continuous variables following the concept of a parameterized action space Markov decision process.
This approach has been used to jointly learn communication and control policies in~\cite{BaumannRL}. Yet, it has only been shown to be successful in low-dimensional tasks.
More recently,~\cite{neunert2020continuous} introduced a hybrid \gls{rl} algorithm that can optimize such problems without reformulation.
So far, this has not been applied to \gls{etc}.
Hierarchical \gls{rl} frameworks~\cite{PrecupPHD,SuttonTDLearning,OptionCriticPaper} represent a third approach to address the hybrid problem setting. Originally, the hierarchical structure stems from the concept of temporal abstraction. However, it also naturally captures the structure of \gls{etc}. 
The discrete high-level decision on which sub-policy to execute next coincides with the communication decision in \gls{etc}, while the sub-policy yields the continuous control command.      
We are not aware of any other work that extends hierarchical \gls{rl} algorithms to make them applicable for \gls{etc}, as we do herein.


\fakepar{Stability analysis} Due to the nonlinear activation functions and many parameters involved, it is usually difficult to provide stability guarantees for learned policies parametrized by \glspl{nn}. Thus, this problem is typically not addressed. Two recent exceptions related to the approach herein are~\cite{bonassi2019lstm} and~\cite{karg2020stability}. In~\cite{bonassi2019lstm}, the authors guarantee input to state stability of long short term memory \glspl{nn}. This can effectively be exploited when using \glspl{nn} for modeling or system identification. Its use for controller design has not yet been demonstrated. 
Karg et al. \cite{karg2020stability} examine the stability of \gls{nn} controllers via output range analysis. This way, they can model the closed-loop behavior of the controlled system and define requirements for asymptotic stability. To achieve this property, they propose to refine the final layer of the \gls{nn}, based on a predefined linear quadratic regulator policy. Our approach is more general in that it refines the entire network instead of only the final layer and is not restricted to a particular type of controllers.
Further, compared to both approaches, we provide results for larger networks and present an approach for checking the stability of \gls{etc} policies, which is generally more challenging due to the sporadic updates, triggered by the communication policy.

\section{Problem Formulation}
\label{sec:ProblemDefinition}



\begin{figure}
\centering
\includegraphics[width=1.0\collen]{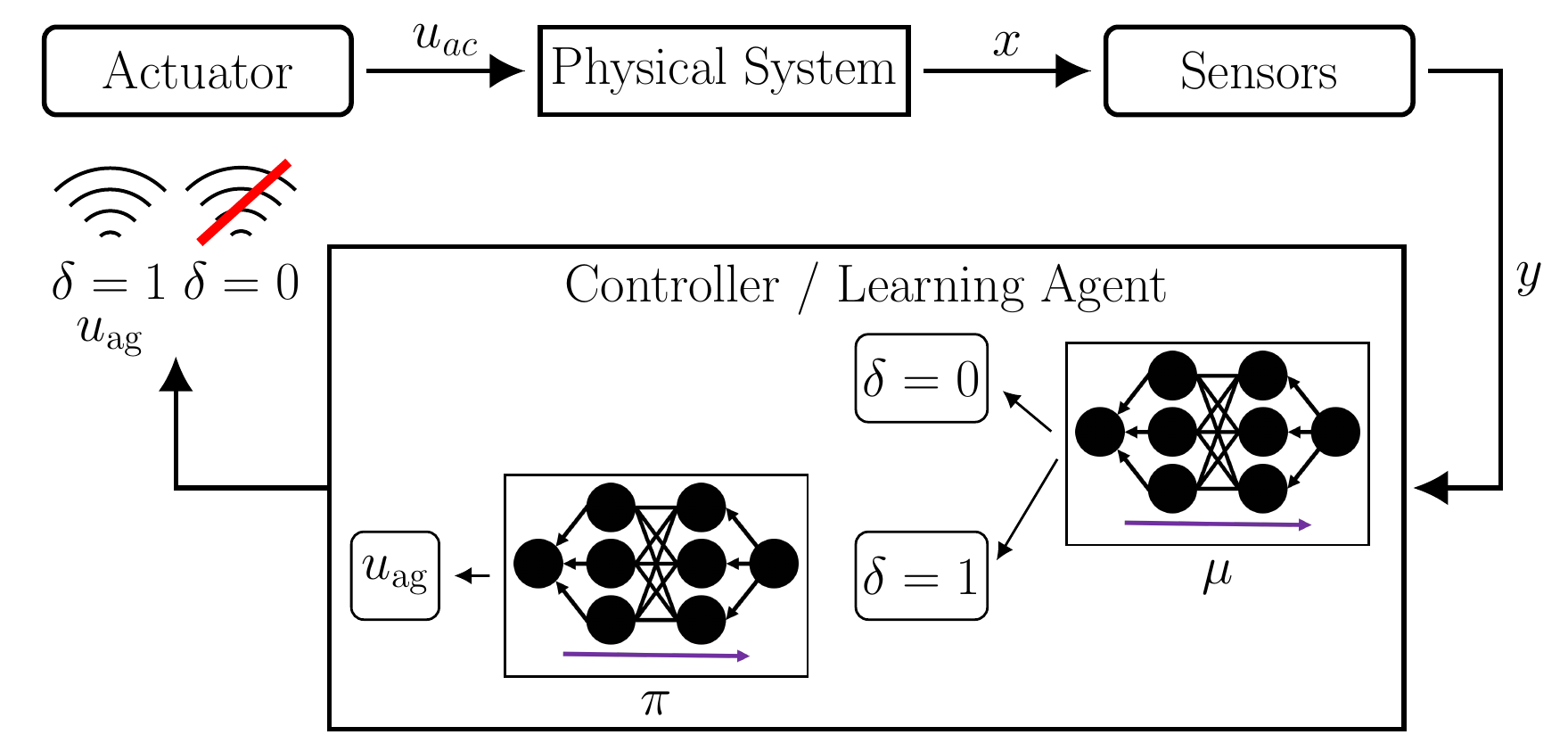}
\caption{Schematic of the learning \gls{etc} setting addressed in this article. An actuator is acting on a physical system. The sensory information about the state of the system is delivered to the learning agent. The agent then has the choice to either communicate a control command to the actuator ($\delta=1$) or to skip the current communication slot to save resources ($\delta=0$). The communication decision is retrieved from \gls{nn} $\mu$. In case of communication, the control action is computed via \gls{nn} $\pi$. During training time, the weights of the \glspl{nn} are refined via backpropagation.}
\label{fig:etc_schematic}
\end{figure}

\fakepar{System} We consider a dynamic system whose state is monitored by sensors, and that is connected to a learning agent (\cf\figref{fig:etc_schematic}). While the sensor measurements are directly available to the learning agent, control commands need to be transmitted over a communication network to the actuators. If not stated differently, we assume the dynamics of the system to be unknown and of the form 
\begin{equation}
x[k+1]=f(x[k],u[k],v[k]) \text{ ,}
\end{equation}
where $x[k] \in \mathbb{R}^n$ denotes the state, $u[k] \in \mathbb{R}^m$ the input, $v[k] \in \mathbb{R}^n$ process noise, and $k \in \mathbb{N}$ the discrete-time index. The measurements $y[k] \in \mathbb{R}^p$ are assumed to be given by 
\begin{equation}
y[k]=g(x[k],w[k]) \text{ ,}
\end{equation}
with $w[k] \in \mathbb{R}^p$ representing measurement noise. For ease of presentation and consistency with existing literature, we assume $g(x,w)=x$ for the derivation of the algorithm in Secs. \ref{sec:hrl} and \ref{sec:leveragingrl_etc1}. However, in the experimental evaluations in Secs. \ref{chp:ResultsHighdimensional} and \ref{chp:HardwareResults}, we demonstrate its applicability in settings with more complex measurement functions.

\fakepar{Event-triggered control} Unlike in periodic control, in \gls{etc} the feedback loop is closed adaptively. Based on the measurements, the agent decides whether ($\delta[k]=1$) or not ($\delta[k]=0$) to communicate with the actuator - \ie whether to exploit the limited resources or to skip the current communication slot. 
As is typically done in \gls{etc}~\cite{heemels2012introduction}, we assume \gls{zoh} at the actuator in between communication events.
Mathematically, this can be expressed as
\begin{equation} \label{eq:zoh_gen}
u_{\text{ac}}[k] = \begin{cases} u_{\text{ag}}[k], & \text{if } \delta[k] = 1 \\
u_{\text{ac}}[k-1], & \text{if } \delta[k] = 0 \text{ ,}
\end{cases}
\end{equation}
where $u_\text{ac}$ denotes the control action applied at the actuator and $u_\text{ag}$ represents the action calculated by the agent (cf. \figref{fig:etc_schematic}). 

\fakepar{Performance objective} The proposed learning algorithm  should result in both a control and communication strategy that ensure that the physical system behaves in the desired way (\eg reaching a desired setpoint) while the resource constraints are taken into account. Both objectives are combined in the following reward function
\begin{equation}
\begin{split}
R &= \sum_{k=0}^N \gamma^k (R_{\text{ctrl}}[k] + R_{\text{comm}}[k])\\ 
&= \sum_{k=0}^N \gamma^k (R_{\text{ctrl}}[k] - \lambda \delta[k]) \text{ ,}
\label{eq:cost_fct}
\end{split}
\end{equation}
where $\lambda$ is the weight on penalizing communication, $\gamma$ the discount factor, and $R_{\text{ctrl}}[k]$ the control reward. If the discount factor $\gamma$ is less than 1, it limits the horizon up to which future rewards are reflected in the value of the function. We express the communication savings $\Gamma$ as the percentage of the available communication slots that have not been used, \ie 
\begin{equation}
\Gamma = 1 - (\sum_{k=1}^N \delta[k])/N
\end{equation}
for a trajectory of length $N$. That is, $\Gamma = \SI{0}{\percent}$ means all communication slots have been used, and $\Gamma = \SI{100}{\percent}$ corresponds to no communication.

\fakepar{Problem statement} 
The goal of this paper is to devise a learning algorithm that maximizes~\eqref{eq:cost_fct}.
To address the drawbacks of existing methods, the algorithm shall meet the following requirements: \emph{(i)} it should be model-free to avoid the need for an accurate dynamics model; \emph{(ii)} it should jointly optimize the control and communication policy since the separation principle does not hold in general for \gls{etc}; \emph{(iii)} it should be applicable to linear and nonlinear, \emph{(iv)} low- and high-dimensional systems.

\fakepar{Towards stability} In addition to learning \gls{etc}, which is the focus of this work, we are also concerned with providing stability guarantees for the policies. This is essential for the practicability of the method, especially considering its potential use in safety critical environments. However, this is particularly difficult when dealing with learned, highly parametric, and nonlinear \gls{nn} control strategies in the \gls{etc} setting. We approach this challenge by defining system stability through control invariant sets, assuming known, linear dynamics, and using \glspl{nn} with \gls{relu} activations. This allows us to devise a second algorithm capable of verifying and refining the \gls{nn} policies to guarantee system stability.






\section{Background: Hierarchical Reinforcement Learning}
\label{sec:hrl}


\gls{etc} represents a hybrid control problem, which is difficult to solve for most standard \gls{rl} algorithms.
In the following, we introduce the options framework established by Precup \etal~\cite{PrecupPHD,OptionCriticPaper}. 
This hierarchical \gls{rl} framework is particularly suitable for learning control policies in hybrid action spaces as it naturally splits the discrete and continuous variables.

To represent the hierarchy, the options framework requires one additional variable, called option $o \in \mathcal{O}$, besides the state $x \in \mathcal{X}$ and the control action $u \in \mathcal{U}$. $\mathcal{O}$, $\mathcal{X}$ and $\mathcal{U}$ represent the set of options, the state, and the action space. To keep notation uncluttered, we omit time dependence in the next two sections and write $x=x[k]$, $x'=x[k+1]$, and similar for other variables. The framework is parametrized by three policies: the policy over options $\mu(o|x)$, the intra option policy $\pi_{o}(u|x)$, and the termination function $\beta_{o}(x)$. The principle in few words: the policy over options decides which option is to be executed. Depending on this choice, the action is sampled from the corresponding intra option policy until the termination function indicates to stop the execution of the current option, which triggers a restart of the procedure. Mathematically, the policy over options $\mu(o|x)$ determines the probability of choosing option $o$, the intra option policy $\pi_{o}(u|x)$ determines the distribution over actions $u$, and the termination function $\beta_{o}(x)$ determines the probability for terminating the execution of option $o$. As shown in \cite{SuttonTDLearning}, this can be rephrased as having a flat action space without the hierarchy,
\begin{equation} \label{eq:conv_flat}
\pi(u|x,o)=(1-\beta_o(x))\pi_o(u|x) + \beta_o(x) \sum_{\tilde{o} \in \mathcal{O}}^{} \mu(\tilde{o}|x) \pi_{\tilde{o}}(u|x) \text{ .}
\end{equation}
After executing action $u$, the next state and option are given by $x'$ and $o'$. 

To optimize the agent's behavior, it is essential to estimate the expected reward of its actions using the Q-function. This estimate directly indicates which actions are more rewarding. In standard \gls{rl}, the Q-function assigns a value to a state-action pair $(x,u)$. In the hierarchical setting, we define it as assigning a value to a state-option pair $(x,o)$,
\begin{equation}
Q(x,o)=  \int\limits_{\tilde{u} \in \mathcal{U}} \pi_{o}(\tilde{u}|x) \hat{Q}(x,o,\tilde{u}) \ \mathrm{d} \tilde{u} \text{ ,}
\label{eq:Q-fct-hierarchicalrl}
\end{equation} 
where $\hat{Q}$ can be computed via $\hat{Q}(x,o,u)=r(x,o,u)+\gamma Q(x',o')$ and $r$ defines the single step reward. In conclusion, the Q-function describes the future reward to be expected when starting from state $x$ and option $o$. $\hat{Q}$ the reward starting from $x$, $o$ and action $u$. 

The options framework aims at maximizing the expected reward, which for a given state $x$ and option $o$ reads
\begin{equation}
\label{eq:reward_def}
\begin{split}
R(x,o) &=[1-\beta_{o}(x)]Q(x,o) \\
&+\beta_{o}(x) \sum_{\tilde{o} \in \mathcal{O}}^{} \mu(\tilde{o}|x) Q(x,\tilde{o}) \text{ .}
\end{split}
\end{equation}
Assuming the termination function $\beta$ is parametrized by parameters $\theta_\beta$, the policy over options $\mu$ by $\theta_\mu$, and the intra option policy $\pi$ by $\theta_\pi$, the gradient of the reward \eqref{eq:reward_def} with respect to the policies is given by
\begin{equation}
\begin{split}
&\frac{\partial R(x,o,\theta_\beta,\theta_\mu)}{\partial \theta_\mu}\\ 
&= \beta_{o}(x,\theta_\beta) \mathbb{E}[\frac{\partial}{\partial \theta_\mu} \log(\mu(o,\theta_\mu|x))Q(x,o)]
\end{split}
\label{eq:vanilla_grad_pol_op}
\end{equation}
and
\begin{equation}
\begin{split}
&\frac{\partial R(x,o,\theta_\beta,\theta_\mu)}{\partial \theta_\beta} =\\
& \frac{\partial \beta_{o}(x,\theta_\beta)}{\partial \theta_\beta} (\sum_{\tilde{o} \in \mathcal{O}}^{} \mu(\tilde{o},\theta_\mu|x) Q(x,\tilde{o}) - Q(x,o)) \text{ .}
\end{split}
\label{eq:vanilla_grad_term}
\end{equation}

For each intra option policy, we seek to maximize
\begin{equation}
Q(x,o)=\int\limits_{\tilde{u} \in \mathcal{U}} \pi_{o}(\tilde{u},\theta_\pi|x)\hat{Q}(x,o,\tilde{u}) \ \mathrm{d}\tilde{u} \text{ .}
\end{equation}
Taking the gradient results in 
\begin{equation}
\frac{\partial Q(x,o,\theta_\pi)}{\partial \theta_\pi} = \mathbb{E}[\frac{\partial}{\partial \theta_\pi} \log(\pi_{o}(u,\theta_\pi|x))\hat{Q}(x,o,u)] \text{ .}
\end{equation}


For continuous action spaces, \cite{OptionCriticContPaper} proposes to use \gls{ppo} \cite{PPOSchulman} for updating the intra option policy. PPO stabilizes the learning process for continuous action tasks by limiting the covariate shift through including a clipping function into the optimization objective $L$. Let $\theta$ denote the parameters to be optimized, and $\pi'$ denote the new policy to be found ($\pi'=\pi(\theta)$), whereas $\pi$ denotes the policy under the old parameters ($\pi=\pi(\theta_\text{old})$) which were used for sampling the state action transitions. The goal of the \gls{ppo} algorithm is to optimize 
\begin{equation}
\begin{split}
L(\pi') &= R(\pi')-R(\pi) \\
&=r(x,o,u^{\pi'})+\gamma Q^{\pi}(x',o') - Q^{\pi}(x,o) \\
&=A^{\pi}(x,o,u^{\pi'}) \approx \frac{\pi_o'(u|x)}{\pi_o(u|x)} A^{\pi}(x,o,u^{\pi}) \text{ .}
\end{split}
\end{equation}
Following the derivations presented in \cite{OptionCriticContPaper}, we have
\begin{equation}
\begin{split}
\frac{\partial L(\theta)}{\partial \theta} &= \mathbb{E} [\frac{\partial}{\partial \theta} \min[\frac{\pi_o'(u|x)}{\pi_o(u|x)} A^{\pi}(x,o,u^{\pi}) , \\ 
&\text{clip}(\frac{\pi_o'(u|x)}{\pi_o(u|x)},1-\epsilon,1+\epsilon) A^{\pi}(x,o,u^{\pi})]] 
\end{split}
\label{eq:ppo_pol_op}
\end{equation}
for the update of the intra option policy, with the advantage $A^{\pi}(x,o,u^{\pi})=r(x,o,u^{\pi})+\gamma Q^{\pi}(x',o')- Q^{\pi}(x,o)$, and the clipping function
\begin{equation}
\text{clip}(a,b,c) = \begin{cases} a, & \text{if } a \in [b,c] \\
b, & \text{if } a < b \\
c, & \text{if } a > c \text{ ,} 
\end{cases}
\end{equation}
assuming $c \geq b$, and $\epsilon$ the range in which no clipping is applied.

Combining \eqref{eq:vanilla_grad_pol_op}, \eqref{eq:vanilla_grad_term}, and \eqref{eq:ppo_pol_op} leads to the \gls{ppoc} framework for continuous action spaces, as presented in \cite{OptionCriticContPaper}. Further, \cite{OptionCriticContPaper} uses \gls{gae} \cite{schulman2015GAE} to calculate $A^{\pi}(x,o,u^{\pi})$. The GAE algorithm allows to tune the bias-variance tradeoff when estimating this advantage term. 


\section[Learning Event-triggered Control and Communication Policies]{Learning Event-triggered Control and \\ Communication Policies}
\label{sec:leveragingrl_etc1}

We now establish the link between the previously introduced hierarchical \gls{rl} algorithm and the \gls{etc} problem formulation and present an algorithm capable of learning \gls{etc} strategies from data through joint optimization. In \secref{sec:mergin} we bridge the gap between ETC and hierarchical RL and show in which ways the problems they solve are related. In \secref{sec:leveragingrl_etc} we first explain why simply applying plain hierarchical \gls{rl} is not enough to solve \gls{etc} problems before we show how existing algorithms and concepts need to be extended to be applicable.

\subsection{Relating Hierarchical Reinforcement Learning to Event-triggered Control}
\label{sec:mergin}

The hierarchical \gls{rl} algorithm presented in the preceding section naturally allows us to represent problems with hybrid action spaces. The policy over options $\mu$ performs a discrete decision (which option to execute next), while the intra option policy $\pi_o$ returns a continuous action, if needed. Thus, in the context of \gls{etc}, the policy over options represents the triggering law, while the intra option policy yields the control action in case of communication.
We always assume that option 0 ($o_0$) corresponds to no communication and to performing the \gls{zoh}, while option 1 ($o_1$) corresponds to sampling a continuous action from a \gls{nn} policy, \ie $u_\text{ag}[k]=\pi_{o_1}(u|x)$ (\cf~\eqref{eq:zoh_gen}).
Thus, the \gls{et} policy saves not only communication, but also computational resources since the intra option policy does not need to be evaluated in case of no communication.
Using this definition of options,~\eqref{eq:zoh_gen} can be rewritten as
\begin{equation} \label{eq:zoh_gen_opt}
u_{\text{ac}}[k] = \begin{cases} u_{\text{ac}}[k-1], & \text{if } \delta[k] = 0 \text{, } o=o_0  \\
u_{\text{ag}}[k]=\pi_{o_1}(u|x), & \text{if } \delta[k] = 1 \text{, } o=o_1 
\text{ .}
\end{cases}
\end{equation}


\subsection{Hierarchical Reinforcement Learning for Event-triggered Control}
\label{sec:leveragingrl_etc}

Applying the concepts introduced in \secref{sec:hrl} is not sufficient for successfully learning \gls{etc} strategies. This is due to the special nature of the \gls{etc} problem. When using the algorithm in a standard, periodic control setting, both options coincide to sampling a continuous action from the respective intra option policy and applying it to the system. On the contrary, in \gls{etc}, choosing option 0, \ie the \gls{zoh}, directly fixes the action and, thus, introduces a great difference between the capabilities of the options. Simply put, in \gls{etc}, the two options cannot compensate for each other as only option 1 is capable of changing the action applied to the system. To account for this difference, we next propose several modifications to the original algorithm, mainly focused on stabilizing the learning process and increasing exploration on the options level. 

Due to the different capabilities of the two options in \gls{etc}, learning the policy over options is very sensitive. However, in the original \gls{ppoc} framework, the authors propose to only use \gls{ppo} for updating the intra option policy, while the policy over options is refined using vanilla policy gradient \eqref{eq:vanilla_grad_pol_op}. In contrast, we propose to also use the \gls{ppo} algorithm for updating the policy over options for two reasons. First, the learning process is more sensitive due to the limited action of the \gls{zoh} in case of no communication. 
Second, as the two policies influence and affect each other, as can be seen in~\eqref{eq:conv_flat}, restricting the update process for both of the policies enhances the overall performance and stabilizes the learning process. Let $\theta_\mu$ denote the parameters of the policy over options, subject to optimization, and $\mu'=\mu(\theta_\mu)$ the new policy, while $\mu=\mu(\theta_{\mu_{old}})$ represents the old one. Using \gls{ppo} to update the policy over options results in
\begin{equation}
\begin{split}
L(\theta_\mu) &= R(\mu')-R(\mu)=Q(x,o(\mu'))-Q(x,o(\mu)) \\
&=r(x,o(\mu'))+\gamma V^{\mu}(x') - V^{\mu}(x)  \\
&= A^{\mu}(x,u^{\mu'}) \approx \frac{\mu'(o|x)}{\mu(o|x)} A^{\mu}(x,o^{\mu}) \text{ ,}
\end{split} 
\end{equation}
where $V$ is the value function and defined as $V(x)=\sum_{\tilde{o} \in \mathcal{O}}^{} \mu(\tilde{o}|x) Q(x,\tilde{o})$. 
Thus, exploiting the \gls{ppo} algorithm, the gradient is given by
\begin{equation}
\begin{split}
\frac{\partial L(\theta_\mu)}{\partial \theta_\mu} &= E [\frac{\partial}{\partial \theta_\mu} \min[\frac{\mu'(o|x)}{\mu(o|x)} A^{\mu}(x,o^{\mu}), \\ &\text{clip}(\frac{\mu'(o|x)}{\mu(o|x)},1-\epsilon,1+\epsilon) A^{\mu}(x,o^{\mu})]] \text{ .}
\end{split}
\end{equation}

For calculating the advantage function in this case, there are two possibilities:
\begin{enumerate}
\item $A^{\mu}(x,o^{\mu}) = Q(x,o) - \max_{\tilde{o} \in \mathcal{O}} Q(x,\tilde{o}) \text{ ;}$
\item $A^{\mu}(x,o^{\mu}) = Q(x,o) - \sum_{\tilde{o} \in \mathcal{O}}^{} \mu_{\text{old}}(\tilde{o}|x) Q(x,\tilde{o}) \text{ .}$
\end{enumerate}
The first possibility represents a greedy approach as the baseline is the maximum Q-value possible, whereas the second one represents the expected Q-value under the old parameters. In preliminary experiments, the first version performed better and is, therefore, used herein.

The discrepancy in the capabilities of the options also affects the exploration process. While any continuous control input can be applied in the case of communication (\ie option 1), option 0 is extremely limited. The algorithm might tend to only choose option 1 as, especially at the beginning of the learning process, the \gls{zoh} will likely result in no improvements of the reward. Hence, we introduce an entropy term in the optimization algorithm. The entropy scheduling ensures that there is enough exploration for the policy over options, which handles the communication decision. The gradient for the policy over options, therefore, equates to 
\begin{equation}
\begin{split}
&\frac{\partial L(\theta_{\mu'})}{\partial \theta_{\mu'}} = \frac{\partial}{\partial \theta_{\mu'}} \mathbb{E} [\min[\frac{\mu'(o|x)}{\mu(o|x)} A^{\mu}(x,o) , \text{clip}(\frac{\mu'(o|x)}{\mu(o|x)}, \\
&1-\epsilon,1+\epsilon) A^{\mu}(x,o)] + \tau \text{log}(\mu'(o|x)) \mu'(o|x)] \text{,}
\end{split}
\end{equation}
where $\tau$ represents the entropy regularization coefficient. Over time, the entropy regularization is reduced as enough exploration has been conducted and, therefore, a rather exploitative behavior is preferred.

Considering the optimization objectives of the original \gls{ppoc} algorithm reveals that the gradients of the policy over options, as well as the termination function, effectively optimize the same objective (see  \eqref{eq:vanilla_grad_pol_op}, \eqref{eq:vanilla_grad_term}). The policy over options is supposed to choose the option with the highest Q-value, and the termination function should terminate the current option when another option has a higher estimated Q-value. As this is the same goal, we remove the termination function (\ie $\beta(x) \equiv 1$). In this new setting, one can also interpret termination as the policy over options deciding to choose another option. This modification simplifies the learning process, as only two policies have to be optimized. It further clearly establishes the link to the \gls{etc} problem formulation. The discrete decision of the policy over options coincides with the triggering law, while the intra option policy represents the control law. 

The resulting algorithm for learning the hierarchical control policy is presented in \algref{alg:learn_etc}. We use \gls{nn} policies to represent the policy over options, the intra option policy, as well as to approximate the Q-function. Details on the network structures are provided in \appref{sec:append_parametrization}.

Due to the \gls{zoh}, we have to include the last control action applied to the system in the state to keep the properties of a Markov decision process. Therefore, we define
$\tilde{x}[k] = (x[k],~u_\text{ac}[k-1])^\transp$.

\begin{algorithm}
\caption{Hierarchical RL for \gls{etc}}
\begin{algorithmic}[1]
  \small
  \State Initialize Clipping $\epsilon$ and Entropy Regularization $\tau$
  \State Inititalize Q-Network $Q(x,o)$
  \State Initialize Policy over options network $\mu(o|x)$
  \State Initialize Intra option policy network $\pi_o(u|x)$
  \For{number of epochs}
  	\State $Q' \leftarrow Q$
  	\State $\mu' \leftarrow \mu$
  	\State $\pi_o' \leftarrow \pi_o$
  	\State Sample $(x,o,u)$-Transitions using current $\mu$, $\pi$
  	\State Use GAE (\cite{schulman2015GAE}) to calculate $A^{\pi}(x,o,u)$
  	\For{number of optimizer iterations}
  		\For{number of options, $o=0,1,2,...$}
  			\State Sample batch
  			\State $A^{\mu}(x,o) = Q'(x,o) - \max_{\tilde{o}} Q'(x,\tilde{o})$
  			\State $L_1(\theta_{\mu'}) = \mathbb{E} [\min[\frac{\mu'(o|x)}{\mu(o|x)} A^{\mu}(x,o) , \text{clip}(\frac{\mu'(o|x)}{\mu(o|x)},$ \phantom a \phantom a \phantom a \phantom a \phantom a \phantom a \phantom a \phantom a \phantom a $1-\epsilon,1+\epsilon) A^{\mu}(x,o)] + \tau \text{log}(\mu'(o|x)) \mu'(o|x)]$ 
  			\State $L_2(\theta_{\pi'}) = \mathbb{E} [\min[\frac{\pi_o'(u|x)}{\pi_o(u|x)} A^{\pi}(x,o,u) , \text{clip}(\frac{\pi_o'(u|x)}{\pi_o(u|x)}, $ \phantom a \phantom a \phantom a \phantom a \phantom a \phantom a \phantom a \phantom . $1-\epsilon,1+\epsilon) A^{\pi}(x,o,u)]]$
  			\State $L_3(\theta_{Q'}) = \mathbb{E} [(Q'-(Q(x,o)+A^{\pi}(x,o,u)))^2]$
  			\State $\theta_{\mu'} \leftarrow \theta_{\mu'} + \alpha_{\theta_\mu} \frac{\partial L_1(\theta_{\mu'})}{\partial \theta_{\mu'}}$
  			\State $\theta_{\pi'} \leftarrow \theta_{\pi'} + \alpha_{\theta_{\pi'}} \frac{\partial L_2(\theta_{\pi'})}{\partial \theta_{\pi'}}$
  			\State $\theta_{Q'} \leftarrow \theta_{Q'} - \alpha_{\theta_{Q'}} \frac{\partial L_3(\theta_{Q'})}{\partial \theta_{Q'}}$
  		\EndFor 
  	\EndFor
  	\Comment{One potential implementation of the entropy scheduling is shown below}
  	\If {epoch \% 1000==0}
  		\State $\tau=\tau /10$
  	\EndIf
  \EndFor
\end{algorithmic}
\label{alg:learn_etc}
\end{algorithm}


\section{Results in Simulation Environments}
\label{chp:ResultsHighdimensional}

To showcase the versatility of the presented algorithm, we apply it to low-dimensional and linear as well as high-dimensional and nonlinear systems.
We first present results for the OpenAI Gym \cite{1606.01540} Pendulum environment.
This simple environment also allows us to compare the algorithm's performance to classical \gls{etc} approaches.
Next, we show its behavior in challenging nonlinear and high-dimensional MuJoCo \cite{todorov2012mujoco} environments.

For all simulation experiments, we use two options as shown in \eqref{eq:zoh_gen_opt}. The reward is given by \eqref{eq:cost_fct}, where, if not stated differently, $R_{\text{ctrl}}[k]$ is the unmodified reward provided by the respective environment. For all the experiments, we use the environments' original sampling rate of \SI{20}{\Hz} and their unmodified measurement functions. Thus, the state available to the learning agent is given by $\tilde{x}[k] = (g(x[k],~w[k]),~u_\text{ac}[k-1])^\transp$.

We train all of the agents for 5000 epochs and store the model that achieves the highest reward. Each epoch consists of sampling 2048 $(x,o,u)$-transitions. For each communication penalty $\lambda$, we start 10 training runs with different seeds and report the performance of the best models. The \gls{nn} architectures are presented in~\appref{sec:append_parametrization}. Training one model took around 30 to 40 hours on a single CPU. However, we did not parallelize the training process, which the algorithm would allow for. Thus, we expect that the training time can be reduced significantly. On a laptop with an Intel® Core™ i7-7700HQ CPU @ 2.80GHz and 24GB RAM, the evaluation of both of the policies (policy over options and intra option policy) takes on average \SI{1.1}{\ms}. The code that has been used to train the learning-based models and videos illustrating the results are available at \footnote{\label{cont}\scriptsize{\url{https://sites.google.com/view/learn-event-triggered-control}}}.

\subsection{Pendulum Environment}
\label{sec:poc}

In this rather simple environment, we consider the challenge of  stabilizing the inverted pendulum on top, already starting in an upright position. We use the reward function's original parameters, except for increasing the penalization on the control input from 0.001 to 0.1 to prevent the controller from being too aggressive.

For this task, it is straightforward to linearize the system dynamics around the equilibrium. This allows us to compare the results of the presented algorithm to other well-known \gls{etc} approaches. In particular, we compare our algorithm to 

{\scriptsize
\begin{itemize}
\item \textbf{LQR}: \\ $u[k] = K x[k] \Rightarrow \delta[k]=1 \text{ } \forall k \text{ ,}$
\item \textbf{LQR random skip}: \\ $u[k] = \begin{cases} K x[k] \Rightarrow \delta[k]=1, & \text{if } \nu>\xi  \\
u[k-1] \Rightarrow \delta[k]=0, & \text{otherwise} \text{ ,} \end{cases} $  
\item \textbf{state triggering 2 norm}: \\ $u[k] = \begin{cases} K x[k] \Rightarrow \delta[k]=1, & \text{if } \norm{x[k]}_2>\xi    \\
u[k-1] \Rightarrow \delta[k]=0, & \text{otherwise} \text{ ,} \end{cases} $
\item \textbf{output based triggering} \cite{6069816}: \\ $u[k] = \begin{cases} K x[k] \Rightarrow \delta[k]=1, & \text{if } \norm{K\hat{x}[k]-Kx[k]}_2>\xi \norm{Kx[k]}_2    \\
u[k-1] \Rightarrow \delta[k]=0, & \text{otherwise} \text{ ,} \end{cases} $
\item \textbf{state diff triggering} \cite{tabuada2007event}: \\ $u[k] = \begin{cases} K x[k] \Rightarrow \delta[k]=1, & \text{if } \norm{\hat{x}[k]-x[k]}_2>\xi \norm{x[k]}_2    \\
u[k-1] \Rightarrow \delta[k]=0, & \text{otherwise} \text{ ,} \end{cases} $
\end{itemize}
}
where $\hat{x}[k]$ represents the state at the last triggering instance and $\xi$ a threshold variable, adjusting the triggering condition. The random variable $\nu$ is uniformly sampled from the interval $[0,1]$. The gain matrix K is chosen from an LQR design where the weights are identical to the parameters of the reward function.

\begin{figure}
\centering
\includegraphics[width=\collen]{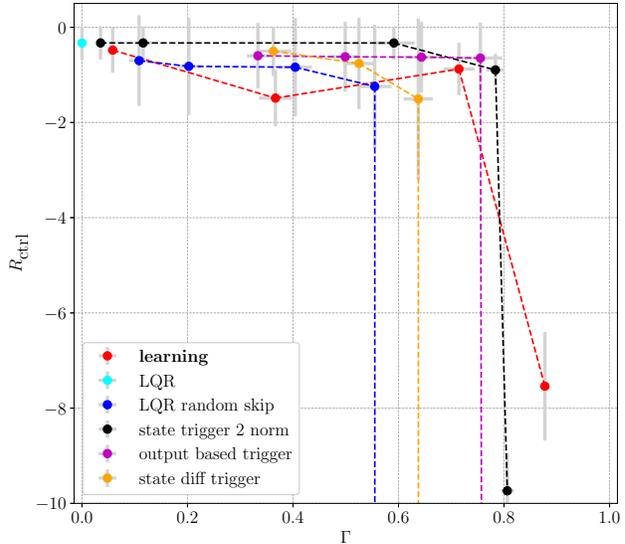}
\caption{Illustration of the magnitude of the control reward $R_\text{ctrl}$ versus the communication savings $\Gamma$ for various \gls{etc} strategies and our learning approach for the task of stabilizing an inverted pendulum. The mean and standard deviation are obtained by performing 10 rollouts with each of the policies, and indicated by the grey lines. The vertical drops indicate when the policies become unstable, \ie fail to stabilize the pendulum. For intermediate communication savings, all the approaches show similar performance. Considering the maximum savings possible, the learning approach outperforms the others.}
\label{fig:pareto_pend_top}
\end{figure}

In \figref{fig:pareto_pend_top}, we show the performance of the introduced event-triggering laws and our data-based \gls{rl} algorithm. As can be seen, for the classical approaches, only communication savings up to about \SI{80}{\percent} are possible, whereas our algorithm finds policies that can save up to \SI{90}{\percent}. However, for intermediate communication savings, the classical methods usually outperform our learning-based approach. This might be caused by the algorithm getting stuck in local optima. Nevertheless, the rewards of the classical approaches and our algorithm are still in the same order of magnitude for the intermediate savings. This distinguishes the herein presented algorithm from the \gls{rl} framework proposed in~\cite{BaumannRL}, which also achieved communication savings of up to \SI{90}{\percent} in the pendulum environment, but was outperformed by orders of magnitudes for intermediate communication savings.

\subsection{Results in high-dimensional, nonlinear Environments}
\label{ref:highdim}

\begin{figure}
\begin{subfigure}{1.0\collen}
	\centering
	\begin{subfigure}[t]{0.475\textwidth}
	\centering
	\includegraphics[width=0.9\textwidth]{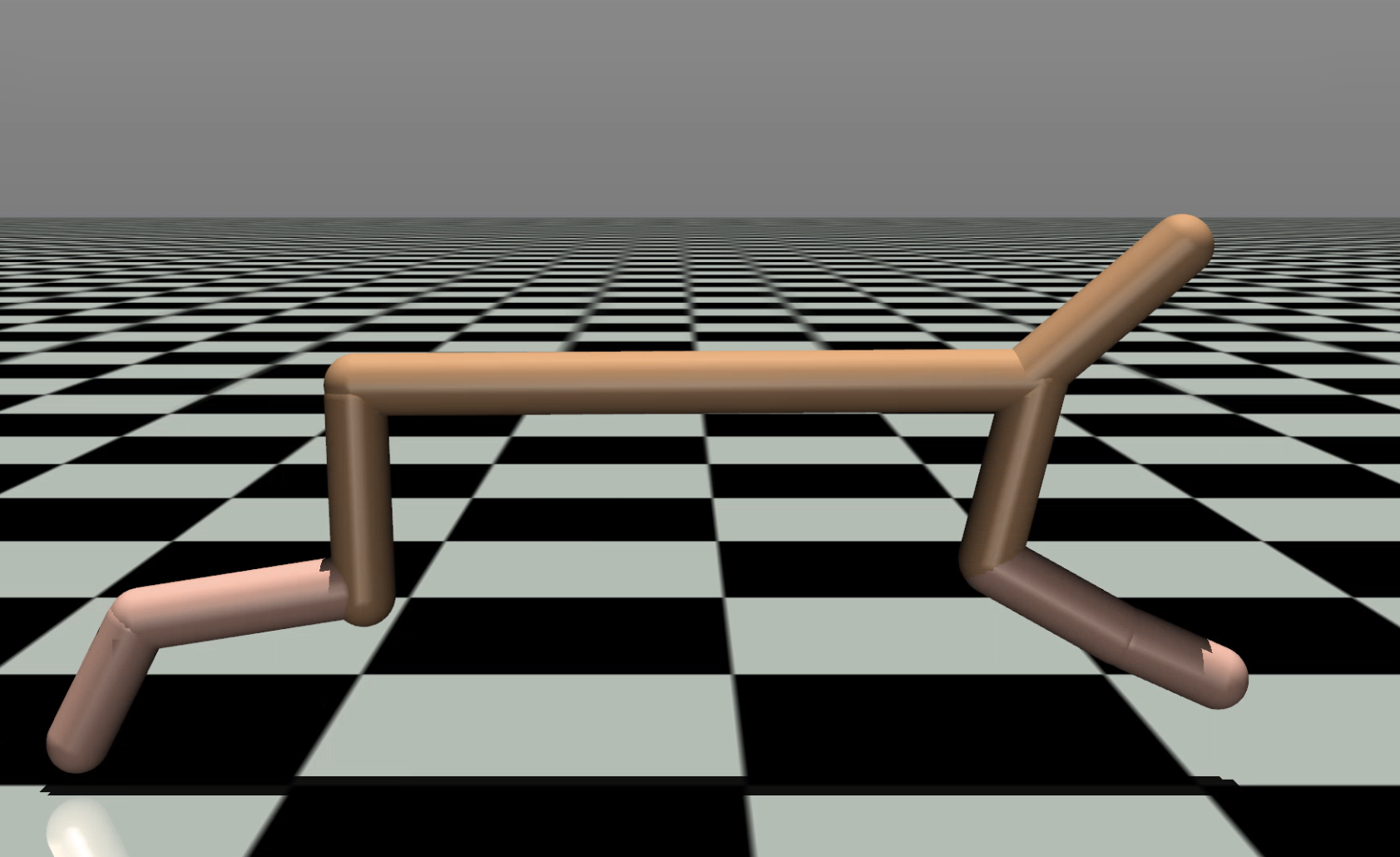}
	\caption{Half-Cheetah environment. Its observation space has 18 dimensions, and the action space 6.}
	\label{fig:cheetah_env}
	\end{subfigure} \hfill
	\begin{subfigure}[t]{0.475\textwidth}
	\centering
	\includegraphics[width=0.9\textwidth]{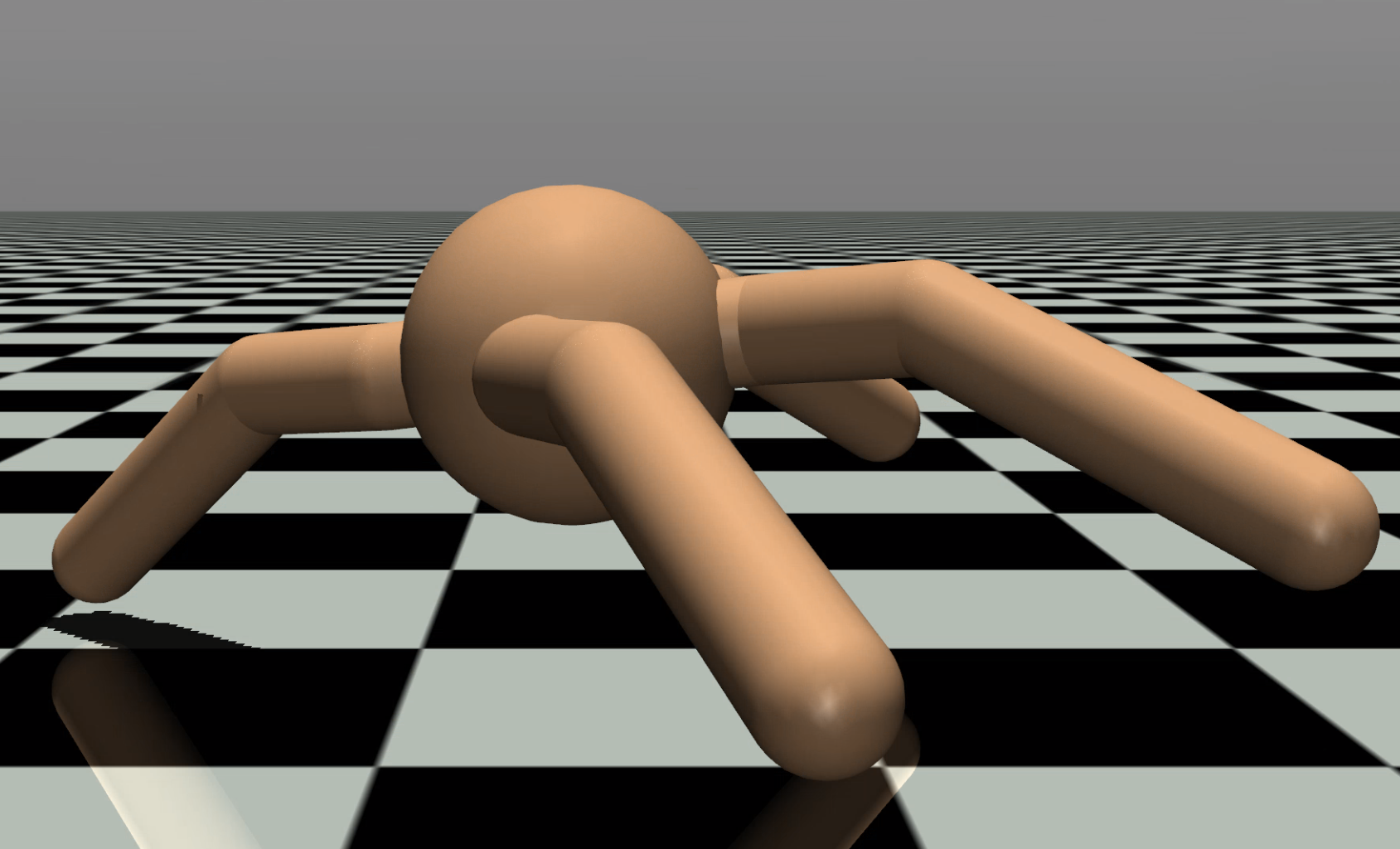}
	\caption{Ant environment. Its observation space has 111 dimensions, and the action space 8.}
	\label{fig:ant_env}
	\end{subfigure} 
\end{subfigure}
\caption{The high-dimensional MuJoCo environments in which we train the learning agents.}
\label{fig:mujoco_envs}
\end{figure}

We now focus on the nonlinear, high-dimensional MuJoCo Half-Cheetah (\figref{fig:cheetah_env}) and Ant (\figref{fig:ant_env}) environment, that cannot simply be linearized around certain equilibria points to yield linear system dynamics. In those challenging nonlinear environments, known approaches for \gls{etc} usually fail as the settings are too complex. To our knowledge, only \cite{BaumannRL} presents results in such environments. However, through separately learning control and communication. 

In these environments, the reward $R_\text{ctrl}$ is mainly made up of the distance covered, but additionally includes a cost on the control action and contact forces. The goal of the optimization is to move the robotic agent as far as possible in the available time, given input and environmental constraints. Apart from the reward, we will also analyze the distance $d$ that the robotic agents cover during a rollout. We slightly modify the Ant environment by eliminating the restriction that limits the jump height of the center of mass of the Ant.

\subsubsection{Half-Cheetah Environment}

\begin{figure}
\centering
\includegraphics[width=1.0\collen]{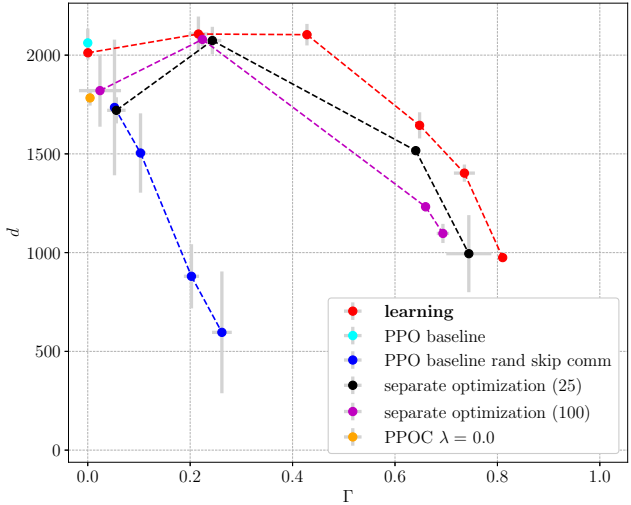}
\caption{Illustration of the performance of various agents. Depicted is the distance traveled by the Half-Cheetah $d$ versus the communication savings $\Gamma$. The shown mean and standard deviation are calculated from the nine best, out of ten rollouts, and illustrated in grey. The red line connects the results obtained from agents, trained using our algorithm. The blue line combines the rollouts of a PPO baseline policy, which saves communication by randomly skipping communication. The purple and black lines connect the results obtained from agents, trained using a modified version of our proposed algorithm. I.e., control and communication are optimized separately, in an alternating fashion. For the black line, the optimization is switched every 25 epochs; for the purple one, every 100 epochs. The herein presented joint optimization approach outperforms the others in terms of performance at intermediate communication savings as well as maximum savings possible.}
\label{fig:pareto_cheetah}
\end{figure}

The red graph in \figref{fig:pareto_cheetah} illustrates the performance of policies trained using the presented algorithm. By increasing the value of $\lambda$, higher communication savings $\Gamma$ can be achieved while, on the other hand, the covered distance decreases. In this Half-Cheetah environment, we are capable of learning policies that can achieve up to \SI{80}{\percent} communication savings. In the figure, we showcase the performance of the nine best, out of ten rollouts with each policy. This is due to the fact that some of the policies exhibit one rollout where the Cheetah is flipped onto its back, resulting in a significantly decreased distance covered.

Interestingly, the communication savings have an impact on the gait, as shown in the accompanying video. More communication savings result in a rather jumpy policy; when the Cheetah is in the air, there is no need to communicate. One can see that for a low penalty on communication, the Cheetah's feet stay rather close to the ground, which also results in slightly faster progress. As shown in the red graph of \figref{fig:pareto_cheetah}, the standard deviation of the distance traveled is small, indicating that all the policies are robust and reliable. The training stability, \ie the percentage of training runs that result in moving the Cheetah forward, is dependent on the communication penalization $\lambda$. The percentage of successful training runs typically decreases as the value of $\lambda$ is increased, as the option of never communicating becomes more and more attractive. Therefore, for high values of $\lambda$, the training runs naturally converge to policies that do not communicate at all. This results in the Cheetah standing still at its initial position.


We also compare our algorithm's results with a baseline policy trained using the PPO algorithm. This baseline PPO policy only learns the control policy, while the communication strategy is fixed to always communicate. One possibility to achieve communication savings using this baseline policy is to randomly skip communication with a predefined probability. As can be seen in the blue graph in \figref{fig:pareto_cheetah} and the corresponding video, this rapidly decreases the performance. 
This emphasizes that it is crucial to optimize the control and communication strategy jointly. The models trained using our algorithm outperform the baselines. They are considerably more resource-efficient while the Cheetah still covers at least the same distance.

As an additional benchmark, we implemented a modified version of the proposed algorithm that optimizes the control and communication strategy separately, in an alternating fashion. The performance of these agents is visualized through the black and purple graph in \figref{fig:pareto_cheetah}. To obtain the agents corresponding to the black and purple line, we switch between solely optimizing control and solely optimizing communication, every 25 respectively 100 epochs. As can be seen, the proposed joint optimization approach outperforms the separate optimization method. Considering low communication savings, the difference in performance is small. However, for high communication savings, it is significant. This highlights that especially for more difficult tasks, where it is crucial to finely adjust the two policies to each other, the joint optimization approach is superior. The experiments also illustrate that a higher frequency of alternation (black graph) results in better performance. Further increasing the frequency of alternation naturally converges to the joint optimization approach.

We additionally investigate the significance of the modifications detailed in \secref{sec:leveragingrl_etc}.
For this, we try to learn an \gls{et} controller using the original \gls{ppoc} implementation (see \secref{sec:hrl}). As shown in \figref{fig:pareto_cheetah}, for $\lambda=0.0$, the \gls{ppoc} algorithm finds a solution that saves a minimal amount of communication. However, for higher values of $\lambda$, e.g., 1.0, the \gls{ppoc} algorithm always results in a policy that never communicates, and therefore, the Cheetah does not progress at all. Thus, unlike our algorithm, the original \gls{ppoc} algorithm is incapable of arriving at \gls{et} controllers that reduce communication significantly, while still moving the Cheetah forward. This is probably due to the greedy optimization of the \gls{ppoc} algorithm. Further insights are provided in \appref{sec:hierarchical_period_ctrl}.

\subsubsection{Ant Environment}

\begin{figure}
\centering
\includegraphics[width=1.0\collen]{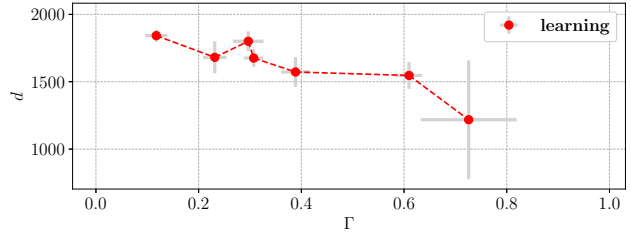}
\caption{Illustration of the agents' performance, trained using our proposed algorithm. Depicted is the distance covered by the Ant $d$ versus the communication savings. The shown mean and standard deviation are calculated from the nine best, out of ten rollouts, and illustrated in grey. At maximum, \SI{70}{\percent} of communication can be saved.}
\label{fig:pareto_ant}
\end{figure}

When deploying the algorithm to the even higher-dimensional Ant environment, we obtain similar results.
\figref{fig:pareto_ant} shows the distance covered by the Ant versus the communication savings $\Gamma$ for models trained using the proposed algorithm. We again show the performance of the nine best, out of ten rollouts for each policy since some policies exhibit one rollout where the Ant is flipped onto its back, resulting in a significantly decreased distance covered. As can be seen from the plot, the most resource-aware control policies are capable of saving up to \SI{70}{\percent} of communication. The figure also illustrates that performance degrades if the communication savings increase. The policy saving \SI{70}{\percent} exhibits the largest standard deviation. Therefore, it is the least stable, but also the most resource-efficient policy. The corresponding video from rollouts with different penalizations underlines those findings and aligns with the results for the Half-Cheetah. When more communication is possible, the Ant's feet are kept rather close to the ground, which results in fast progress, and more reliability as the chances of flipping are minimized. When the penalization on communication is progressively increased, the gait changes towards a rather jumpy behavior. This allows for the most significant communication savings, as when the feet are in the air, no communication is needed. Nevertheless, compared to the results for the Cheetah, the changes in the gait behavior are less obvious. Considering the training process, again, if the resource constraints become very restrictive, the learning process becomes more difficult, and never communicating also becomes a local optimum. 

Compared to \cite{BaumannRL}, where the authors also try to learn an \gls{etc} strategy from data for the Ant environment, the proposed joint optimization achieves even higher communication savings with up to \SI{70}{\percent}, whereas~\cite{BaumannRL} only reported savings of up to \SI{60}{\percent} for their separate optimization approach.


\section{Results on Hardware}
\label{chp:HardwareResults}

While many deep \gls{rl} methods have shown good performance in simulation environments, there are only few examples where the learned policies are actually deployed on real hardware. 
This is mainly due to the sample inefficiency of those algorithms or insufficient simulation to reality transfer.
In this section, we present the results of using our learning algorithm on real hardware, namely, the Apollo robot shown in \figref{fig:hw_apollo_cup}. 

In contrast to the previous simulation experiments, which provide perfect communication and no delays, we now consider real experimental conditions with noise, delays, and potential packet losses. 
The experimental setup is as follows: the sensory information, \ie Apollo's joint configuration, is communicated to the learning agent running on a computer. The agent then decides whether or not to communicate, and, depending on that decision, eventually sends new control commands to the robot. We use a base sampling time of $\SI{20}{\hertz}$. Further details on the setup can be found in \appref{sec:supp_material_robot_exp}.

\subsection{Problem Definition and Setup}
\label{sec:apollo_setup_definition}

To demonstrate the feasibility of the introduced framework, we employ a real-time position controller for the end-effector of Apollo's right arm with 6 \gls{dof}. The controller that is to be learned operates in the cartesian space. The goal is to reach a desired position as accurately as possible under resource constraints, \ie trying to communicate as efficiently as possible. The learning agent's output is a desired reference velocity in the task space of the end-effector,
$u[k]=v_{\text{ref}}[k]=(v_x[k],~v_y[k],~v_z[k])^\transp$.

As the input to the robot has to be with respect to the individual joints, while the learning algorithm's output is defined in task space, the procedure is as follows. The learning algorithm outputs velocities in task space, which are then mapped to the corresponding commands in joint space through inverse kinematics. We use the Pinocchio library~\cite{carpentier2019pinocchio} for this step. The process of computing the target velocity in joint space is iterative and affected by the current robot configuration. It is not guaranteed that the desired velocity can be reached accurately. Once the velocities in joint space are computed, they are applied until the next instance of communication. In the beginning, the desired velocity is reached quite accurately. Considering a longer horizon, it is obvious that constant joint velocities do not coincide with a constant velocity in task space. Therefore, over time, the discrepancy between the desired and the actual velocity increases as there is no recomputation until the next instance of communication.

Since the learning agent only operates in the task space, the state which is fed to the agent is 
$\tilde{x}[k] = (g(x[k],w[k]),~u_\text{ac}[k-1])^\transp =  (\dot{x}_\text{ef}[k],~x_{\text{ef}}[k],~ x_{\text{ref}}[k] - x_{\text{ef}}[k],~u_\text{ac}[k-1])^\transp$, 
where $x_\text{ef}$ denotes end-effector position and $x_\text{ref}$ denotes the desired reference position that is to be reached, in cartesian coordinates. Thus, no joint information is available to the learning agent, which makes the setting partially observable.

In the settings in \secref{ref:highdim}, the steady-state solution is applying a sequence of control inputs to achieve a constant movement of the agent. However, for this task, once the end-effector is close to the reference, it is desirable to apply a zero action to hold its position. As it is unlikely that a \gls{nn} policy outputs exactly zero, we exploit our approach's hierarchical nature and simply define a third option ($o_2$) that corresponds to setting the control input to zero. Hence, for the hardware experiments, unlike presented in \eqref{eq:zoh_gen_opt}, we consider three options: 
\begin{equation} \label{eq:hw_opt}
u_{\text{ac}}[k] = \begin{cases} u_{\text{ac}}[k-1], & \text{if } \delta[k] = 0 \text{, } o=o_0  \\
u_{\text{ag}}[k]=\pi_{o_1}(u|x), & \text{if } \delta[k] = 1 \text{, } o=o_1 \\
 u_{\text{ag}}[k]=\begin{pmatrix}0 & 0 & 0 \end{pmatrix}^\transp, & \text{if } \delta[k] = 1 \text{, } o=o_2
\text{ .}
\end{cases}
\end{equation}

To obtain the results presented in the next sections, we first train the learning algorithm in simulation using the simulation laboratory (SL) framework \cite{SL__2009}. Then, we deploy the learned policies on the real robot. For the transfer from simulation to reality, no additional adjustments are performed.

\subsection{Dynamic Reference Position Tracking}
\label{sec:dynamic_reaching}

\begin{figure}
\centering
\includegraphics[width=0.6\collen]{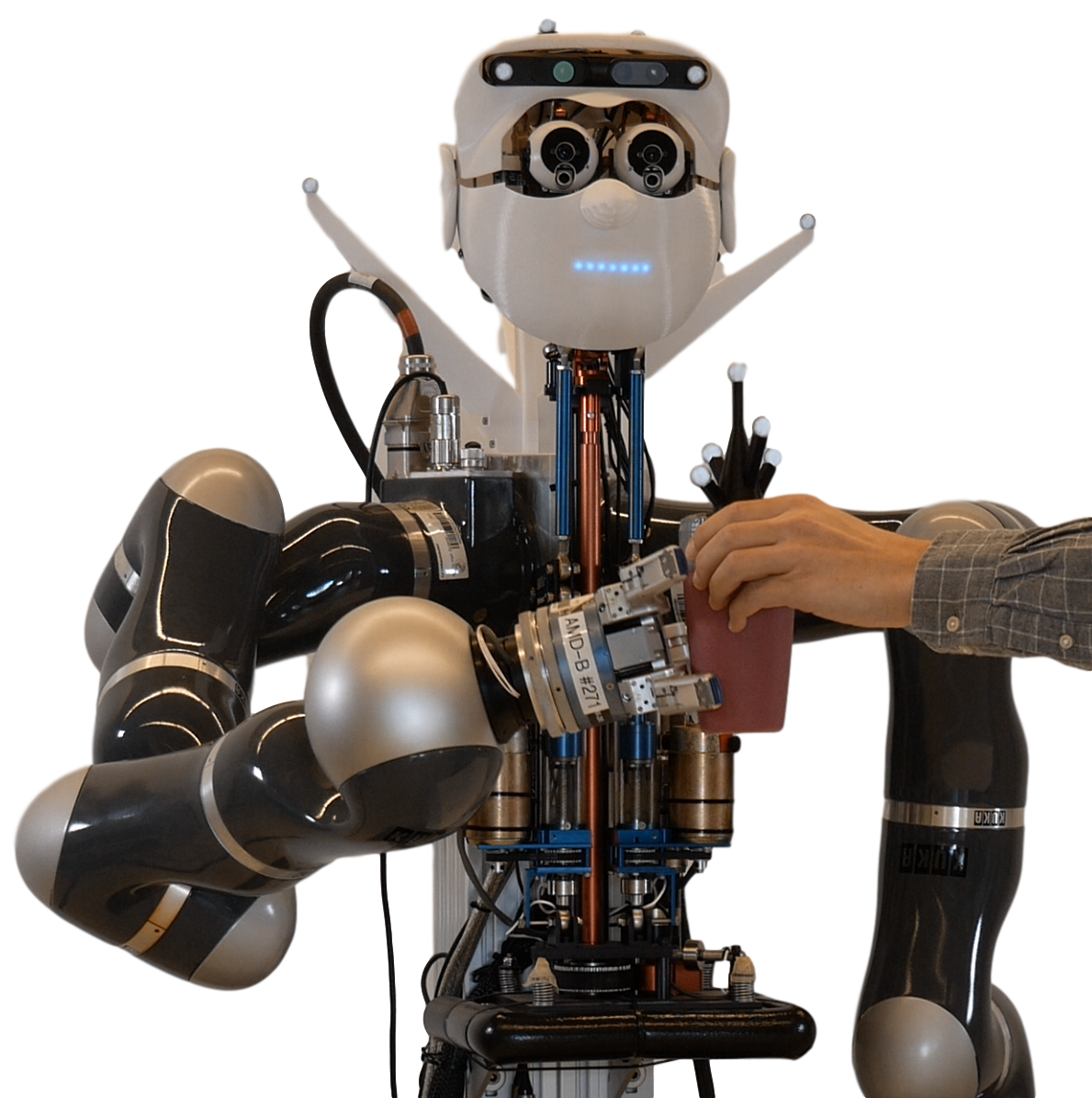}
\caption{Illustration of the hardware experiment where the goal is to place Apollo's hand close to the cup. The cup's position is estimated using the Vicon system.}
\label{fig:hw_apollo_cup}
\end{figure}

At first, we describe the performance of resource-aware agents trained for reaching a dynamic reference position. I.e., putting the end-effector close to a cup (\cf \figref{fig:hw_apollo_cup}) such that it could be grabbed. The cup's position is estimated using the Vicon system and included in the sensory information.

During training, we choose a random initial position of the robot arm and a random reference position (representing the cup) inside the interval 
$([-0.2, 0.5],~[0.55, 0.9],~[-0.15, 0.45])^\transp\SI{}{\meter}$, 
for each trajectory. At each simulated timestep, the reference position is reset to another randomly sampled point inside this interval with probability \SI{1}{\percent}. This procedure should already account for the fact that during the evaluation on the real system, the reference will change dynamically. 

The policies, whose results are presented in the following, have been trained for \num{2250} epochs in simulation with a communication penalty of $\lambda=0.1$.
To incentivize reaching the final position more accurately, we added an inverse term to the reward function $R$, which is thus given by $R = \sum_{k=0}^N \gamma^k (R_{\text{ctrl}}[k] + R_{\text{comm}}[k]) = \sum_{k=0}^N \gamma^k (-3(0.01 {\norm{u[k]}_2}^2 + 10 {\norm{x_{\text{ref}}[k] - x_{\text{ef}}[k]}_2}^2 + 0.01 {\norm{\dot{x}_\text{ef}[k]}_2}^2   + \lambda \delta[k]) + \frac{0.05}{{\norm{x_{\text{ref}}[k] - x_{\text{ef}}[k]}_2}^2})$. The parameters have been obtained empirically and reflect a standard cost function with an increased emphasis on reaching the target while ensuring that the overall cost stays in the same order of magnitude as for the previous experiments. We chose a factor of $0.05$ for the inverse term as we view \SI{5}{\cm} as close enough to the goal position.

\begin{figure}
\centering
\includegraphics[height=0.9\textheight,keepaspectratio,width=\collen]{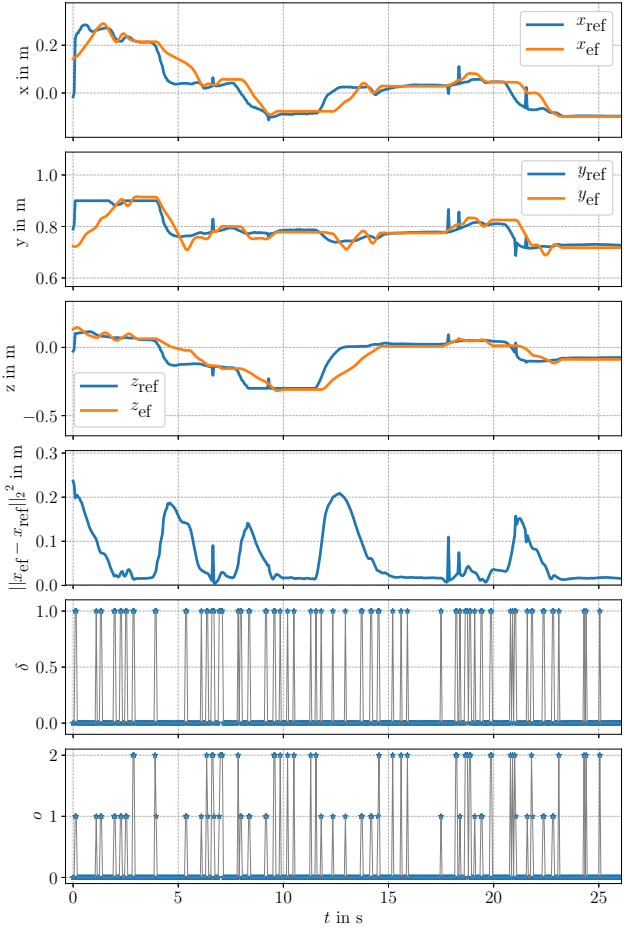}
\caption{Exemplary rollout of a policy trained for the task of reaching a dynamic reference position with Apollo's right hand, see \figref{fig:hw_apollo_cup}. The reference (\ie the cup's position) is provided by the Vicon system. During the rollout, the policy saves $85\%$ of communication, while the reference is still tracked reliably. This performance is consistent among various runs.}
\label{fig:hw_rollout_dyn}
\end{figure}

In \figref{fig:hw_rollout_dyn}, we depict an exemplary rollout of the model while tracking the reference signal provided by the Vicon system. It is striking that $85\%$ of communication can be saved while still reaching the dynamic desired reference reliably. As can be seen, the reference signal is sometimes slightly disturbed. However, although only trained in simulation without such nonidealities, the learning agent is robust to these disturbances. Thus, we conclude that we did not overfit and that our problem formulation and implementation results in a stable and robust simulation to reality transfer.
The corresponding video illustrates the responsiveness of the final policy and confirms the low tracking error. The learned controller drives the hand close enough to the cup such that it can be grabbed.

\subsection{Obstacle Avoidance}

To further illustrate our learning algorithm's capabilities, we increase the difficulty and demonstrate the resulting performance for the inherently more complex, nonlinear task of reaching a desired end-effector position in the face of an obstacle, as illustrated in \figref{fig:title_hw_apollo}. Now, it is not sufficient to simply drive the end-effector in the direction of the reference. Instead, multiple changes of direction are necessary to avoid the obstacle.

The state available to the agent is the same as introduced in \secref{sec:apollo_setup_definition}. Thus, the learning agent is only aware of the end-effector position $x_\text{ef}$ in task space. The algorithm's goal is to reach the desired reference position $x_\text{ref}$ without hitting the obstacle with the end-effector. We assume a static obstacle position, as otherwise, this information would need to be provided to the agent. The reference position is now defined in software and not retrieved via the Vicon system. 

\begin{figure}
\centering
\includegraphics[height=0.9\textheight,keepaspectratio,width=\collen]{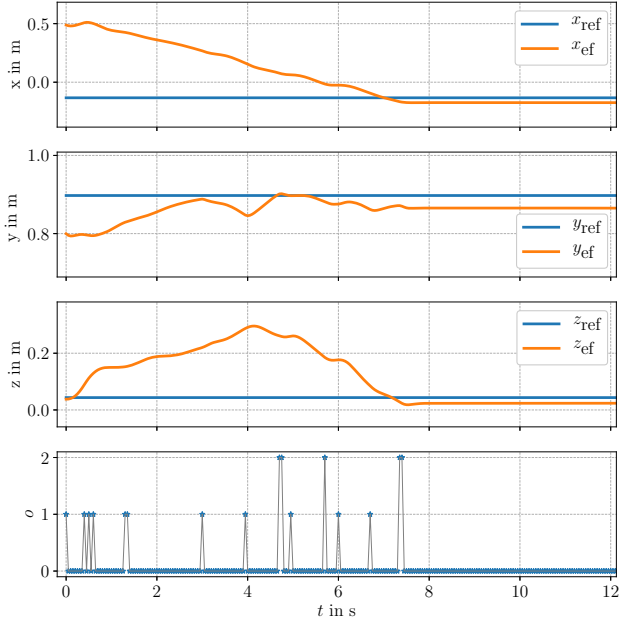}
\caption{Exemplary obstacle avoidance trajectory, where a fixed reference position is to be reached in the face of an obstacle, cf. \figref{fig:title_hw_apollo}. During the rollout, the policy saves \SI{92}{\percent} of communication, while the reference is reached reliably without hitting the obstacle. This performance is consistent among various runs.}
\label{fig:hw_rollout_obst}
\end{figure}

We adapt the previously presented training procedure as follows. The starting and reference positions are sampled from the intervals 
$([0.45, 0.55],~[0.7, 0.9],~[0.0, 0.2])^\transp \SI{}{\meter}$, 
and 
$([-0.2, -0.1],~[0.7, 0.9],~[-0.1, 0.1])^\transp \SI{}{\meter}$,
respectively. This way, they are separated, with the obstacle in between at 
$([-0.05, 0.35],~[0.5, 1.0],~[-1.0, 0.1])^\transp \SI{}{\meter}$.
Moreover, during training only, before applying the control action to the system, we predict the next position of the end-effector using integrator dynamics, \ie $x_\text{pred,ef}[k+1]=x_\text{ef}[k]+\Delta T u_\text{ac}[k]$. If the next predicted position $x_\text{pred,ef}[k+1]$ lies inside the obstacle, we apply 
$u_\text{ac}[k]= (0,~0,~0)^\transp$ 
to the system, and additionally penalize this state/action combination in the reward function. Thus, the reward is now given by $R = \sum_{k=0}^N \gamma^k (R_{\text{ctrl}}[k] + R_{\text{comm}}[k] + R_{\text{obst}}[k]) = \sum_{k=0}^N \gamma^k (-3(0.01 {\norm{u[k]}_2}^2 + 10 {\norm{x_{\text{ref}}[k] - x_{\text{ef}}[k]}_2}^2 + 0.01 {\norm{\dot{x}_\text{ef}[k]}_2}^2   + \lambda \delta[k]) + \frac{0.05}{{\norm{x_{\text{ref}}[k] - x_{\text{ef}}[k]}_2}^2} - \zeta[k])$, where
\begin{equation*}
\zeta[k] = \begin{cases} 0, & \text{if } x_\text{pred,ef}[k+1] \text{ not inside obstacle area} \\
5, & \text{otherwise.}
\end{cases}
\end{equation*}
For obtaining the results presented in \figref{fig:hw_rollout_obst}, we first pretrain the policy for 2250 epochs in simulation and then evaluate it on the real hardware.

As shown in \figref{fig:hw_rollout_obst}, \figref{fig:title_hw_apollo}, and the associated video, the reference position is reached reliably without hitting the obstacle, while still \SI{92}{\percent} of communication can be saved. This behavior is consistent among different starting and reference positions. However, compared to the previously presented reference tracking, without the obstacle present, the reference position is not reached as accurately. The reason for this behavior is that the presence of the obstacle limits the freedom of the arm and the policy, resulting in less accuracy. The fact that the reference position is now defined in software and not noisy could explain why more communication can be saved compared to the previously presented dynamic cup reaching scenario (\secref{sec:dynamic_reaching}).

\section{Towards Stability of Event-triggered Control}
\label{sec:stability_statment}

In the previous sections, we tackle the \gls{etc} problem formulation using model-free, deep \gls{rl}. The results of our learning algorithm are \gls{nn} policies consisting of many parameters. In this setting, it is usually difficult to provide stability guarantees. Yet, since the learned policies are also envisioned to be used in real-world scenarios, such guarantees are essential. In the following, we present an approach toward checking the stability of learned \gls{nn} policies.
As a starting point, in this work, we restrict to \glspl{nn} parametrized with \glspl{relu}, and known, linear system dynamics.

\subsection{Stability Verification of Neural Network Policies}
\label{sec:stabi_veri_nn}

The following stability analysis is based on the Marabou framework proposed in~\cite{katz2017reluplex,katz2019marabou}, which allows to check for properties of deep \glspl{nn}. The Marabou framework is an expansion of the Simplex method \cite{dantziglinear}. Simplex tries to find a valid assignment to a linear program. It either returns an admissible value for all the free variables involved that result in satisfying the query, or returns that the query is not satisfiable. As piecewise linear activation functions can be interpreted as case dependent linear constraints, the authors expand the algorithm such that the same satisfiability queries, as for the Simplex method, can be posed with respect to \glspl{nn}.

We exploit this framework to come up with stability guarantees for trained policies. For our considerations, the only limitation is that exclusively piecewise linear activation functions, \ie \gls{relu} activations can be used within the networks. Nevertheless, by combining the algorithm with assumptions on the system dynamics, it is possible to provide stability guarantees through output range analysis for the learned \gls{nn} policies.
Further, in case the policy does not fulfill the stability requirements straight away, the framework can be used to develop a retraining procedure to refine the \glspl{nn}. 


\subsection{Stability Analysis and Retraining of Event-triggered Control Policies}

We define stability as finding a positive invariant set in the state space: once inside this set, when applying the \gls{nn} policy, the next state is guaranteed to also lie within this set. 

\begin{defi}[Stability]{$\text{ }$\\}
The system $f$ is considered stable under the \gls{nn} control policy~$u_\text{NN}[k]=h_\text{NN}(\tilde{x}[k])$ if there exists a region $\mathcal{M}$, such that \\
$x[k] \in \mathcal{M} \Rightarrow x[k+1]\!=\!f(x[k],h_\text{NN}(\tilde{x}[k])) \in  \mathcal{M} \text{ } \forall x[k] \in\mathcal{M}$.
\label{def:stability_def}
\end{defi}

While the \gls{rl} algorithm presented in \algref{alg:learn_etc} is model-free and can be used for systems with linear or nonlinear dynamics, the stability check relies on the underlying system exhibiting known, linear dynamics. The algorithm then checks whether the control policy is stable, according to \defref{def:stability_def}. Linear system dynamics can also be represented using \glspl{relu}. Thus, algorithmically, we can design one \gls{nn} that takes as the input the current state $x[k]$ and outputs the next state $x[k+1]$, \cf \figref{fig:MarabouSchematicET}. With this network given, we can exploit the Marabou framework \cite{katz2019marabou} to check for the desired properties. 

\subsubsection{Stability Analysis of the Policies}

\begin{figure}
\centering
\includegraphics[width=1.0\collen]{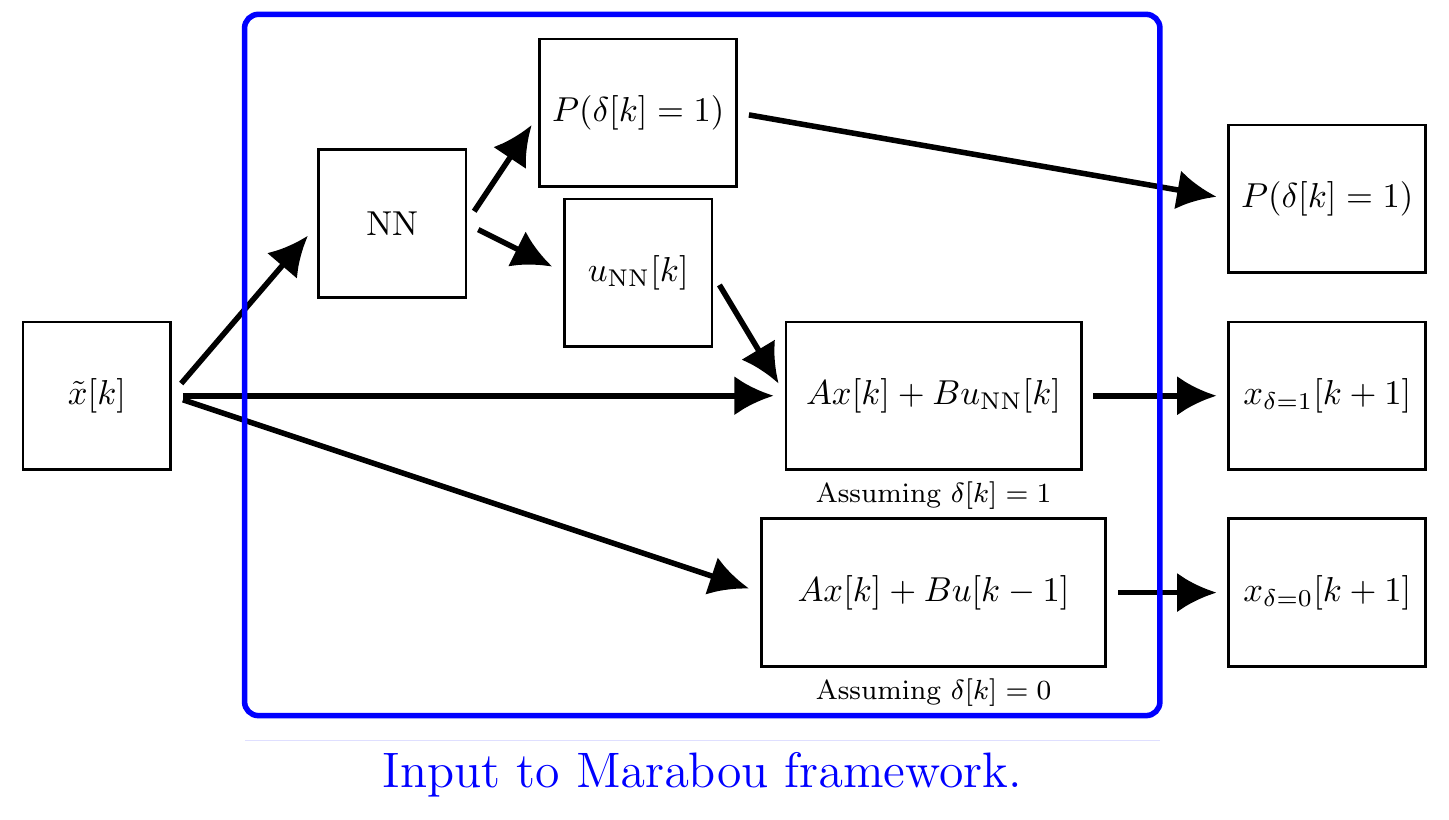}
\caption{Schematic of the pipeline used for the stability verification procedure. The parts inside the blue box can all be parametrized using \gls{relu} units. Therefore, the Marabou framework can check whether the next state is inside the same set as the initial state. This is the basis for our stability analysis.}
\label{fig:MarabouSchematicET}
\end{figure}

In the \gls{et} setting, the next state depends on the communication decision. Therefore, effectively two next states have to be calculated, as shown in \figref{fig:MarabouSchematicET}. 
To make use of the deterministic Marabou framework, we need to eliminate the stochasticity of the communication decision.
Hence, we always choose to communicate in case the probability is larger than or equal to \SI{50}{\percent}.

The resulting algorithm that checks for the stability of a region $\mathcal{M}$ is presented in \algref{alg:check_stab_et}. As the input to the stability verification framework (see \figref{fig:MarabouSchematicET}) is given by $\tilde{x}[k]$, we additionally define 
$\mathcal{S} = (\mathcal{M},~\mathcal{L})^\transp = (\mathcal{M},~[-u_\text{lim},u_\text{lim}])^\transp$, 
which combines the stable region with the input range. Adding the input range is necessary as it covers the potential range of the previously applied control actions, which are reapplied in case the \gls{zoh} is selected. If the algorithm returns an empty set of points, we know that given the region $\mathcal{M}$, the input range, and the system dynamics, the \gls{etc} policy is stable. Otherwise, the algorithm directly outputs exemplary unstable points $\tilde{x}[k]$.  
There are two possibilities, why a point $\tilde{x}[k]$ is unstable under the \gls{etc} policy. Either the policy erroneously decides to skip communication, and the reapplication of the previous control input results in the next state outside the invariant set. Or, in case of communication, the policy chooses an unstable control input.

\begin{algorithm}
\caption{Check for Stability in ET setting ($\mathcal{S}$)}
\label{alg:check_stab_et}
\begin{algorithmic}[1]
  \scriptsize
  \State points=$[]$
  \State Marabou query: $\tilde{x}[k] \in \mathcal{S} \text{, } P(\delta[k]=1)\geq 0.5 \Rightarrow x_{\delta=1}[k+1] \in  \mathbb{R}^{n} \setminus \mathcal{M}$
  \If {Valid assignment is found:} {points.append($\tilde{x}[k]$)} \EndIf
   \State Marabou query: $\tilde{x}[k] \in \mathcal{S} \text{, } P(\delta[k]=1) < 0.5 \Rightarrow x_{\delta=0}[k+1] \in  \mathbb{R}^{n} \setminus \mathcal{M}$
  \If {Valid assignment is found:} {points.append($\tilde{x}[k]$)} 
  \EndIf
  \State \textbf{return} points
\end{algorithmic}
\end{algorithm}

\subsubsection{Retraining Neural Network Policies}

If \algref{alg:check_stab_et} does not indicate stability, \ie returns a non-empty set of points, it is possible to refine the \gls{etc} policy such that it fulfills the invariance property using \algref{alg:refine_pol ET}.

\begin{figure}
\centering
\includegraphics[width=0.6756\collen]{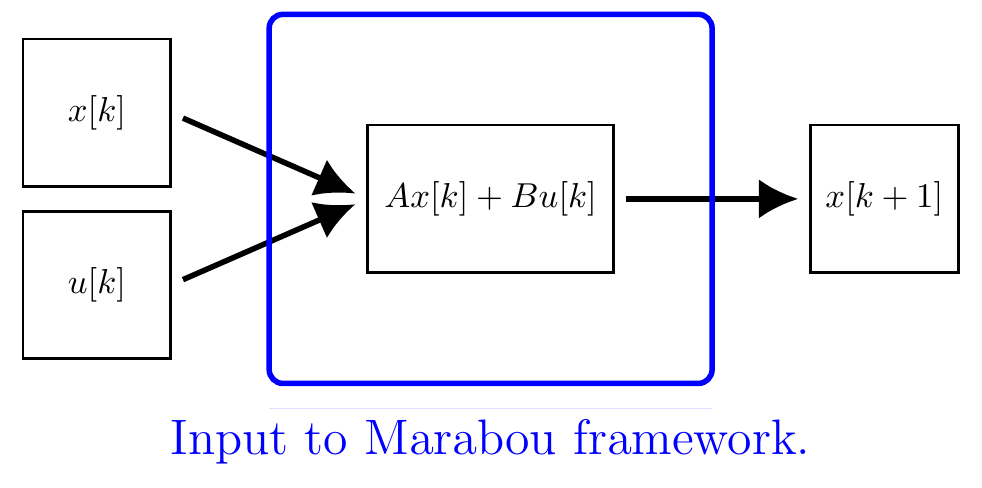}
\caption{Schematic of the pipeline that is used to find an admissible input $u[k]$ for an unstable point $x[k]$. The parts inside the blue box can all be parametrized using \gls{relu} units.}
\label{fig:InputCheckSchematic}
\end{figure}

At the core of this algorithm is the \textit{FindValidInputET} function (ll.~\ref{line:start_validinput_et}-\ref{line:end_validinput_et}). In case the stability verification algorithm (\algref{alg:check_stab_et}) returns an unstable point, we feed it into the schematic presented in \figref{fig:InputCheckSchematic} and Marabou returns a value for the input $u[k]$ such that the next state $x[k+1]$ also lies inside the invariant set.

In case a point $\tilde{x}[k]$ is unstable using the current policy, both parts of the \gls{et} controller have to be adapted. On the one hand, it is crucial to ensure that an appropriate control action is chosen. On the other hand, we have to ensure that the policy decides to communicate. This supervised retraining for unstable points can be seen in ll.~\ref{line:start_supervised-line}-\ref{line:end_supervised-line} of \algref{alg:refine_pol ET}. We define that at those points, the probability of communication should be set to \SI{60}{\percent}. Note that any choice above \SI{50}{\percent} is admissible.

When using the previously described mechanism, all the refined policies exhibit significantly reduced communication savings as any unstable configuration can only be resolved using communication. Thus, we found it beneficial to sample additional points using Sobol sequences \cite{sobol1976uniformly} to cover the whole state space as thoroughly as possible (l.~\ref{line:sobol_samp} of \algref{alg:refine_pol ET}). We then add the condition to the retraining procedure that communication must be saved whenever possible (ll.~\ref{line:start_commsav_et}-\ref{line:end_commsav_et} and \ref{line:start_supervised_enforce-line}-\ref{line:end_supervised_enforce-line}). This can also be interpreted as trying to be maximally resource-efficient.

Combining the individual parts results in \algref{alg:refine_pol ET}. 
If the algorithm terminates, we know that the \gls{etc} law is stable for the defined region $\mathcal{M}$, the input range, and the assumed system dynamics.

\begin{algorithm}
\caption{Refine Policy ET}
\label{alg:refine_pol ET}
\begin{algorithmic}[1]
  \scriptsize

  \Function{FindValidInputET}{$\tilde{x}[k],\mathcal{M}$}  \label{line:start_validinput_et}
	\State Marabou query: $\text{find } u[k] \text{, s.t. } x[k], u[k] \Rightarrow x[k+1] \in \mathcal{M}$	  	
  	\State \Return $u[k]$
  \EndFunction \label{line:end_validinput_et}
  \item[]

  \Function{CheckPointET}{$\tilde{x}[k],\mathcal{M}$} \label{line:start_checkpoint_et}
	\State points=$[]$
	\State Marabou query: $\tilde{x}[k] \Rightarrow x[k+1] \in  \mathbb{R}^{n} \setminus \mathcal{M}$ 
  \If {Valid assignment is found:} {points.append($\tilde{x}[k]$)} 
  \EndIf
  \State \textbf{return} points
  \EndFunction \label{line:end_checkpoint_et}
  \item[]
  
  \Function{CommSavingPossible}{$\tilde{x}[k],\mathcal{M}$}  \label{line:start_commsav_et}
  	\State points=$[]$
  	\Comment{Check whether next state is also stable without communicating.}
  	\State Marabou query: $\tilde{x}[k] \Rightarrow x_{\delta=0}[k+1] \in  \mathcal{M} \text{ for } P(\delta[k]=1)$ $\geq 0.5$
  	\If {Valid assignment is found:} {points.append($\tilde{x}[k]$)} 
  	\EndIf
	\State \Return points
  \EndFunction \label{line:end_commsav_et}
  \item[]    
  
  \State Choose region $\mathcal{M}$
  \State Define region $\mathcal{S} = (\mathcal{M},~[-u_\text{lim},u_\text{lim}])^\transp$
  \State $\text{points}_\text{crit}$ $=$ Check for Stability ($\mathcal{S}$), see \algref{alg:check_stab_et}
  \While {not $\text{points}_\text{crit}$ is EMPTY}
	\Comment{Use Sobol sequences to generate additional points}
	\State $\text{points}_\text{sobol}$ $=$ Use sobol sequence to sample from $\mathcal{S}$ \label{line:sobol_samp}
    \State $\text{points}_\text{commsav} = []$
    \For {$\tilde{x}[i]$ in $\text{points}_\text{sobol}$ }
  		\State $\text{points}_\text{crit}$.append(\Call{CheckPointET}{$\tilde{x}[i],\mathcal{M}$}
  		\State $\text{points}_\text{commsav}$.append(\Call{CommSavingPossible}{$\tilde{x}[i],\mathcal{M}$}
  	\EndFor  
  	\Comment{Calculate admissible input for all the unstable points.}
	\State $u_\text{crit}=[]$  	
  	\For {$\tilde{x}[i]$ in $\text{points}_\text{crit}$}
  		\State $u_\text{crit}$.append(\Call{FindValidInputET}{$\tilde{x}[i],\mathcal{M}$}
  	\EndFor
  	\Comment{Supervised retraining to enforce communication savings}
  	\For {number of optimizer epochs} \label{line:start_supervised_enforce-line}
  		\State $L_2(\theta_2)=(P(\delta[k]=1|\text{points}_\text{commsav},\theta_2) - 0.4)^2$
  		\State $\theta_2 = \theta_2 - \alpha_{\theta_2} \frac{\partial L_2}{\partial \theta_2}$
  	\EndFor \label{line:end_supervised_enforce-line}
  	\Comment{Supervised retraining of the NN policy for critical points}
  	\For {number of optimizer epochs} \label{line:start_supervised-line}
  		\State $L_1(\theta_1)=(u_\text{NN}(\text{points}_\text{crit},\theta_1)-u_\text{crit})^2$
  		\State $L_2(\theta_2)=(P(\delta[k]=1|\text{points}_\text{crit},\theta_2) - 0.6)^2$
  		\State $\theta_1 = \theta_1 - \alpha_{\theta_1} \frac{\partial L_1}{\partial \theta_1}$
  		\State $\theta_2 = \theta_2 - \alpha_{\theta_2} \frac{\partial L_2}{\partial \theta_2}$
  	\EndFor \label{line:end_supervised-line}
  \State $\text{points}_\text{crit}$ $=$ Check for Stability ($\mathcal{S}$), see \algref{alg:check_stab_et}
  \EndWhile
\end{algorithmic}
\end{algorithm}

\subsection{Simulation Example}

In the following section, we demonstrate the previously presented algorithm. The task is to obtain a stable \gls{et} controller for stabilizing the inverted pendulum on top in the OpenAI Gym Pendulum environment. To obtain the linear dynamics required for the algorithm, we linearize the nonlinear pendulum dynamics around the upper equilibrium point $\theta=\SI{0}{\degree}$ and $\dot{\theta}=\SI{0}{\degree / \second}$. The conversion from continuous to discrete-time is done using the matrix exponential method. We use the exact same configuration as in \secref{sec:poc}. In line with Definition \ref{def:stability_def}, we define the safe state space ($\mathcal{M}$) to capture ranges of $\theta \in [\SI{-2.5}{\degree},\SI{2.5}{\degree}]$ and $\dot{\theta} \in [\SI{-5}{\degree / \second},\SI{5}{\degree / \second}]$.  Usually, the state of the pendulum environment that is fed into the \glspl{nn} is given by 
$x[k]=(\cos(\theta[k]),~\sin(\theta[k]),~\dot{\theta}[k])^\transp$. 
However, for the stability verification algorithm, we need to express the \gls{nn} policies' output in terms of the system variables, \ie $\theta[k]$ and $\dot{\theta}[k]$. Thus, we can neither use the nonlinear sine nor the cosine function and linearize the state of the environment around the equilibrium, which results in 
$x[k] \approx (1.0,~\theta[k],~\dot{\theta}[k])^\transp$.

\begin{figure}
\begin{subfigure}{1.0\collen}
	\centering
	\begin{subfigure}[t]{0.5225\textwidth}
	\includegraphics[width=\textwidth]{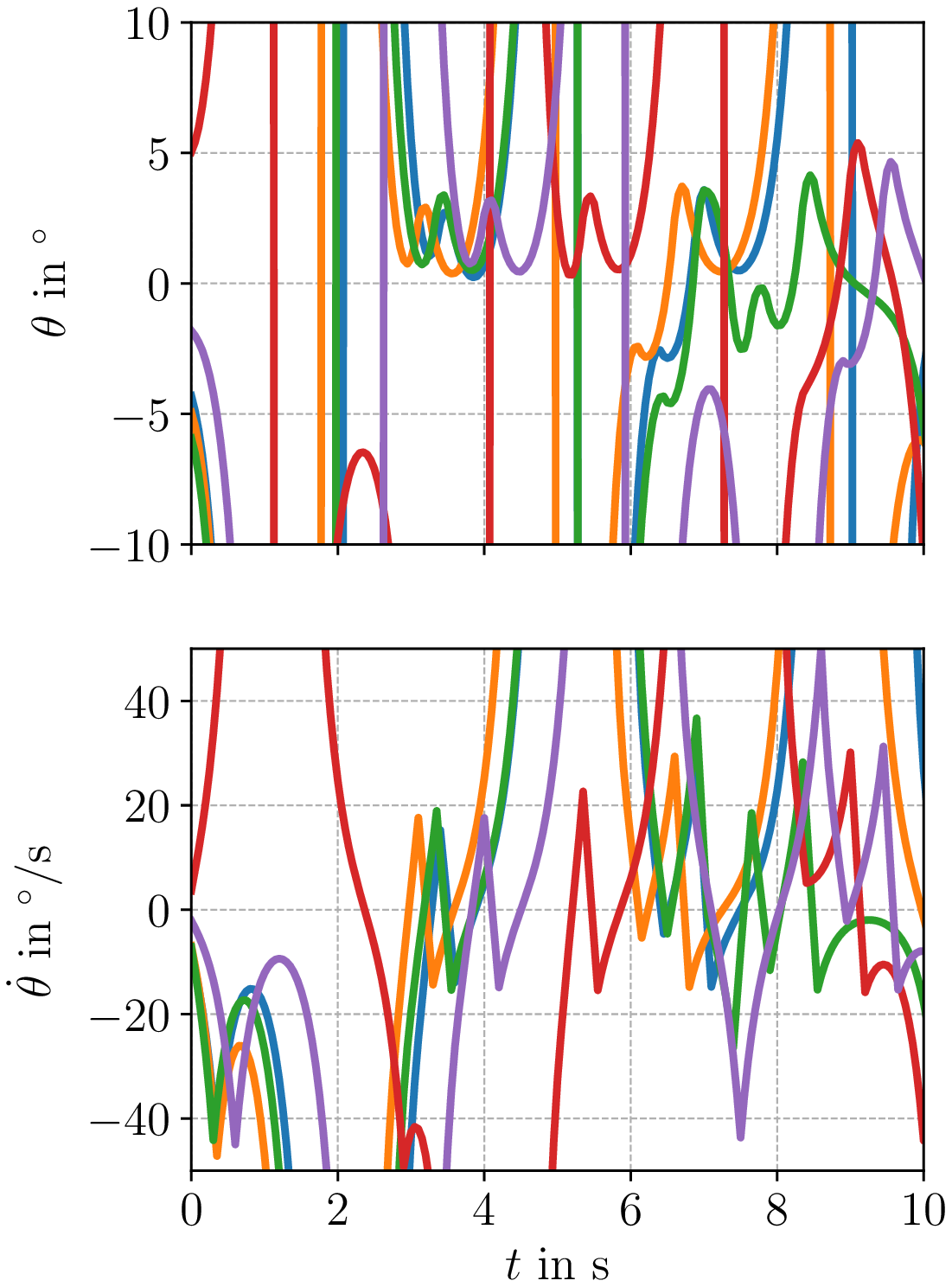}
	\caption{Exemplary rollouts for trying to stabilize a pendulum with an \gls{et} policy, which has been trained for 500 optimization epochs. This policy does not succeed in keeping the pendulum upright.}
	\label{fig:retrain_et_pendulum_unstable}
	\end{subfigure} \hfill
	\begin{subfigure}[t]{0.4275\textwidth}
	\includegraphics[width=\textwidth]{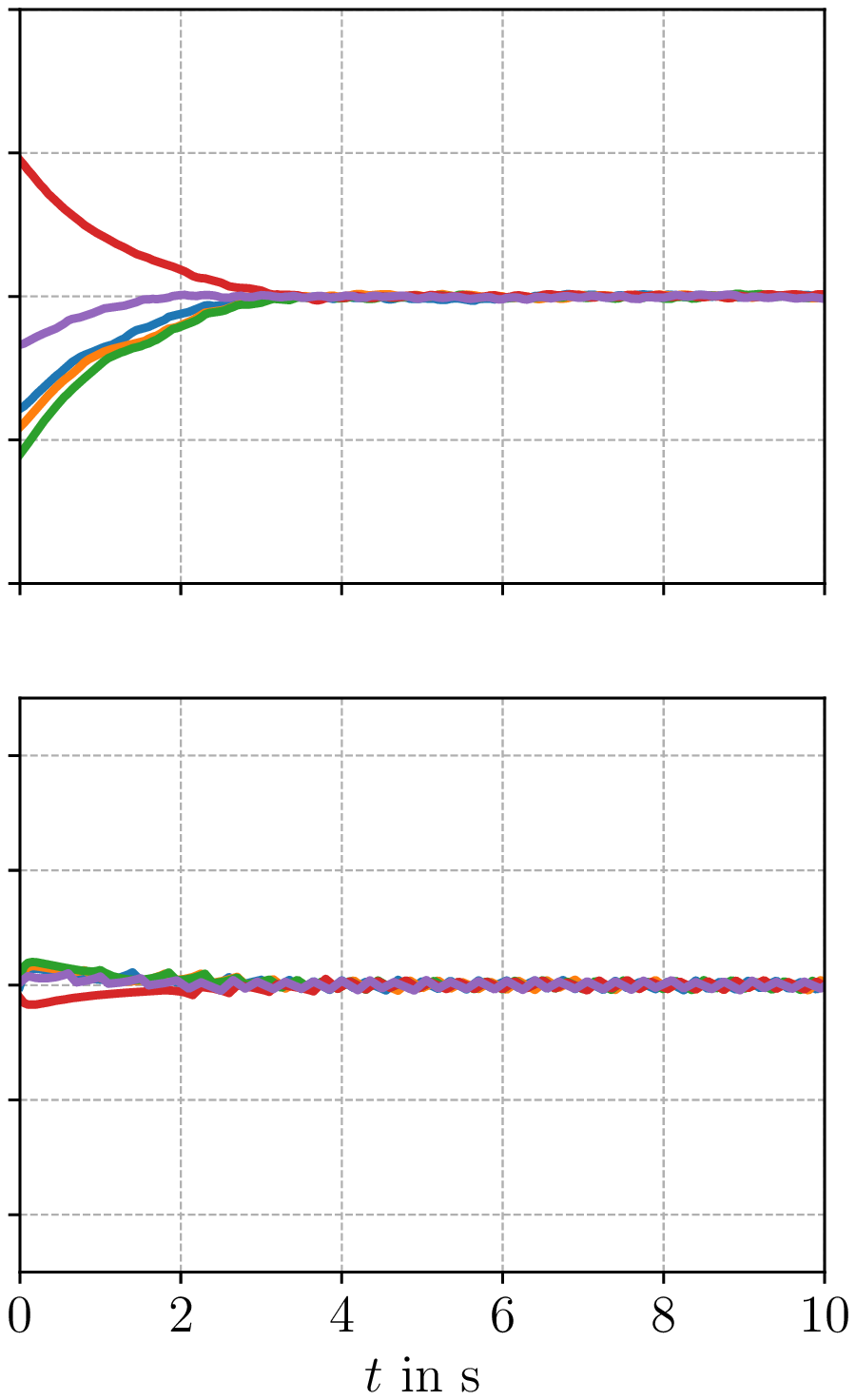}
	\caption{Exemplary rollouts for stabilizing a pendulum after retraining the policy from \figref{fig:retrain_et_pendulum_unstable} with \algref{alg:refine_pol ET} for 4 epochs. This policy is guaranteed to stabilize the pendulum on top.}
	\label{fig:retrain_et_pendulum_stable}
	\end{subfigure}
\end{subfigure}
\caption{Effect of the retraining procedure presented in \algref{alg:refine_pol ET}. In the plots, each color represents a different rollout. While the initial policy is unstable (\figref{fig:retrain_et_pendulum_unstable}), after refining this policy for 4 iterations using \algref{alg:refine_pol ET}, a guaranteed stable controller is obtained which successfully stabilizes the pendulum and still saves about \SI{70}{\percent} of communication (\figref{fig:retrain_et_pendulum_stable}).}
\label{fig:retrain_et_pendulum}
\end{figure}

\begin{figure}
\centering
\includegraphics[width=1.0\collen]{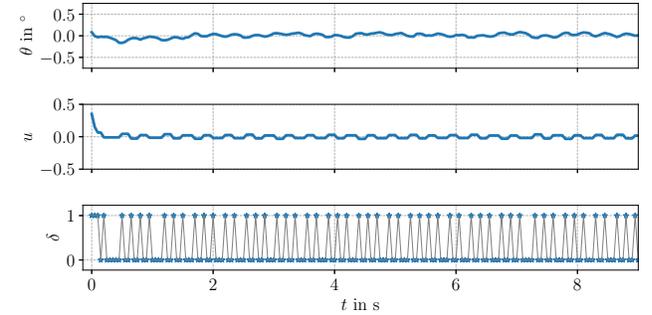}
\caption{Illustration of an exemplary rollout of the same policy as shown in \figref{fig:retrain_et_pendulum_stable}.}
\label{fig:retrain_et_pendulum_rollout}
\end{figure}

The results of running the retraining procedure (\algref{alg:refine_pol ET}) are shown in \figref{fig:retrain_et_pendulum}. We initialize the procedure with the linearized dynamics and a policy that has been trained for 500 epochs using our proposed training algorithm (\algref{alg:learn_etc}). As can be seen in \figref{fig:retrain_et_pendulum_unstable}, this policy is not stable yet. Running the retraining procedure for 4 epochs results in a guaranteed stable \gls{et} controller that also successfully stabilizes the nonlinear system around the equilibrium point (see \figref{fig:retrain_et_pendulum_stable}). Although we only trained for an angular region between \SI{-2.5}{\degree} and \SI{2.5}{\degree}, the policy also keeps the pendulum upright when starting outside of this interval, as shown in \figref{fig:retrain_et_pendulum_stable}. Over 10 randomly started runs, the policy saves around \SI{70}{\percent} of communication. An exemplary rollout of this very resource-efficient and guaranteed stable control policy is presented in \figref{fig:retrain_et_pendulum_rollout}.



\section{Conclusion}
\label{chp:Conclusion_and_Outlook}

In this paper, we propose a model-free hierarchical \gls{rl} algorithm capable of jointly learning event-triggered control policies from scratch. 
Without any modification, the algorithm can be applied to linear and nonlinear, low- and high-dimensional systems.
In those high-dimensional environments, communication savings of up to \SI{80}{\percent} can be reported. To the best of our knowledge, our algorithm is the first that can obtain event-triggered policies for such environments through joint optimization.

The algorithm is also successfully deployed on real hardware, \ie the Apollo robot. We provide a demonstration for resource-efficient setpoint tracking and obstacle avoidance while saving around \SI{85}{\percent} and \SI{90}{\percent} of communication, respectively. These results imply that the presented algorithm scales to partially observable settings, to using more than 2 options, and to imperfect communication settings with potential delays.

Moreover, we show a novel algorithm for evaluating the stability of linear systems controlled by \gls{nn} policies. In case the learned \gls{et} policy initially does not yield the desired stability guarantee, we propose a retraining procedure for refining the previously unstable policy. 
Scaling those ideas to nonlinear environments, as well as higher-dimensional systems through handling the increased computational complexity, is subject to ongoing research.

\begin{ack}                               
The authors would like to thank S. Heim and F. Grimminger for helpful discussions and input. This work was supported in part by the German Research Foundation within the SPP 1914 (grant TR 1433/1-1), the Cyber Valley Initiative, and the Max Planck Society.  
\end{ack}

\bibliographystyle{unsrt} 
\bibliography{bibliography}           

\appendix

\section{Parametrization of the Proposed Algorithm}
\label{sec:append_parametrization}

In this section, we provide insights on how we implement the major components needed for the proposed learning algorithm (\algref{alg:learn_etc}). 

\begin{figure}
\centering
\includegraphics[width=1.0\collen]{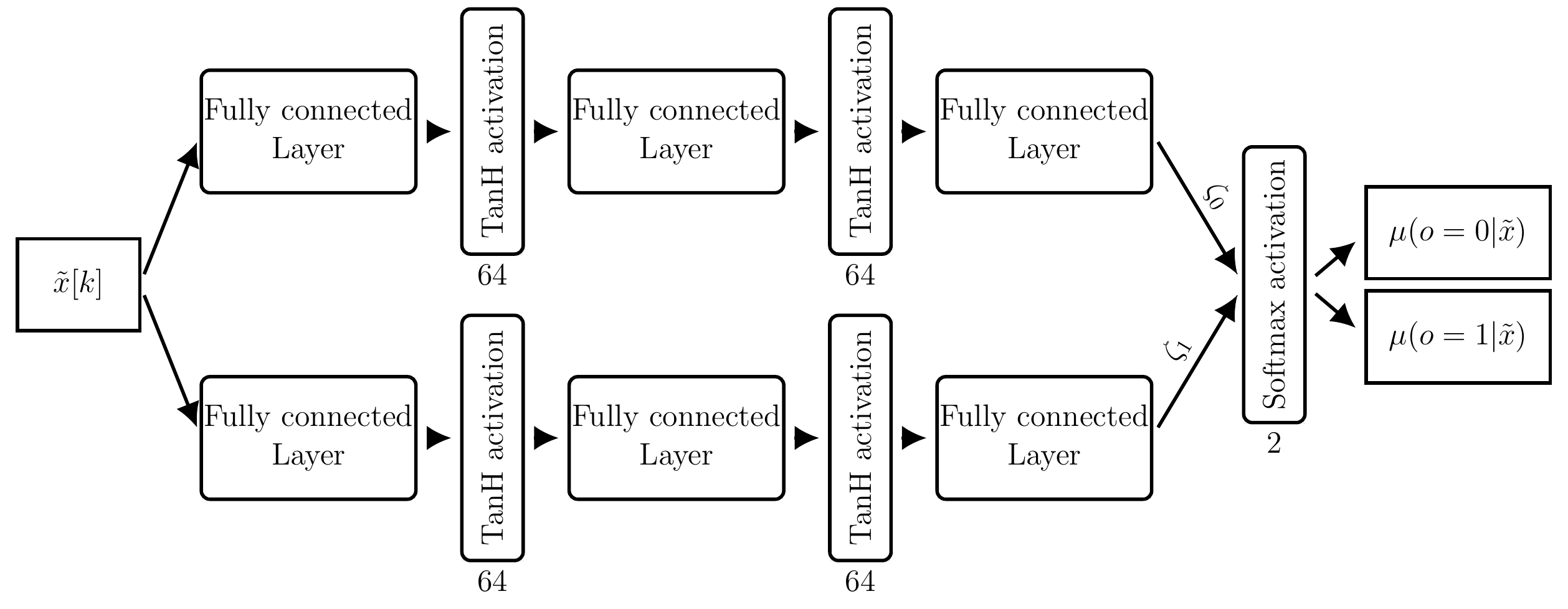}
\caption{Illustration of the parametrization of the policy over options. In this specific implementation, each hidden layer consists of 64 neurons, and the \gls{tanh} and softmax activation functions are used. The variables $\zeta_0$ and $\zeta_1$ represent intermediate values.}
\label{fig:arch_pol_op}
\end{figure}

\begin{figure}
\centering
\includegraphics[width=1.0\collen]{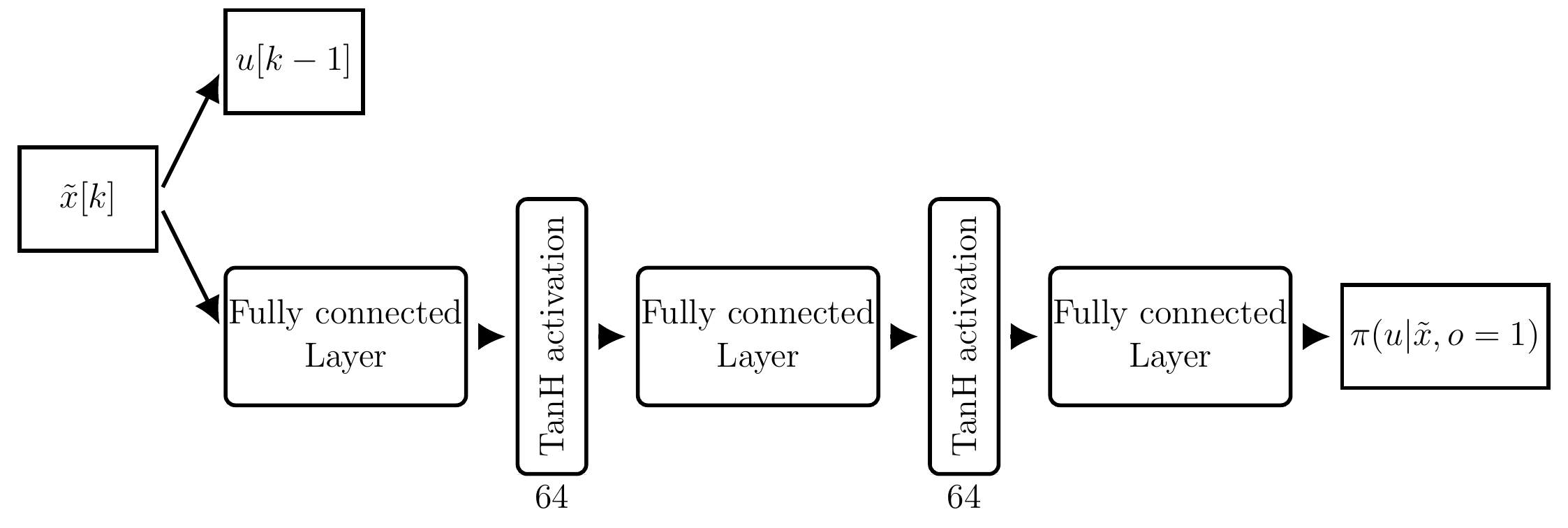}
\caption{Illustration of the parametrization of the intra option policy. In this specific implementation, each hidden layer consists of 64 neurons, and the \gls{tanh} activation function is used.}
\label{fig:arch_intra_op}
\end{figure}

\begin{figure}
\centering
\includegraphics[width=1.0\collen]{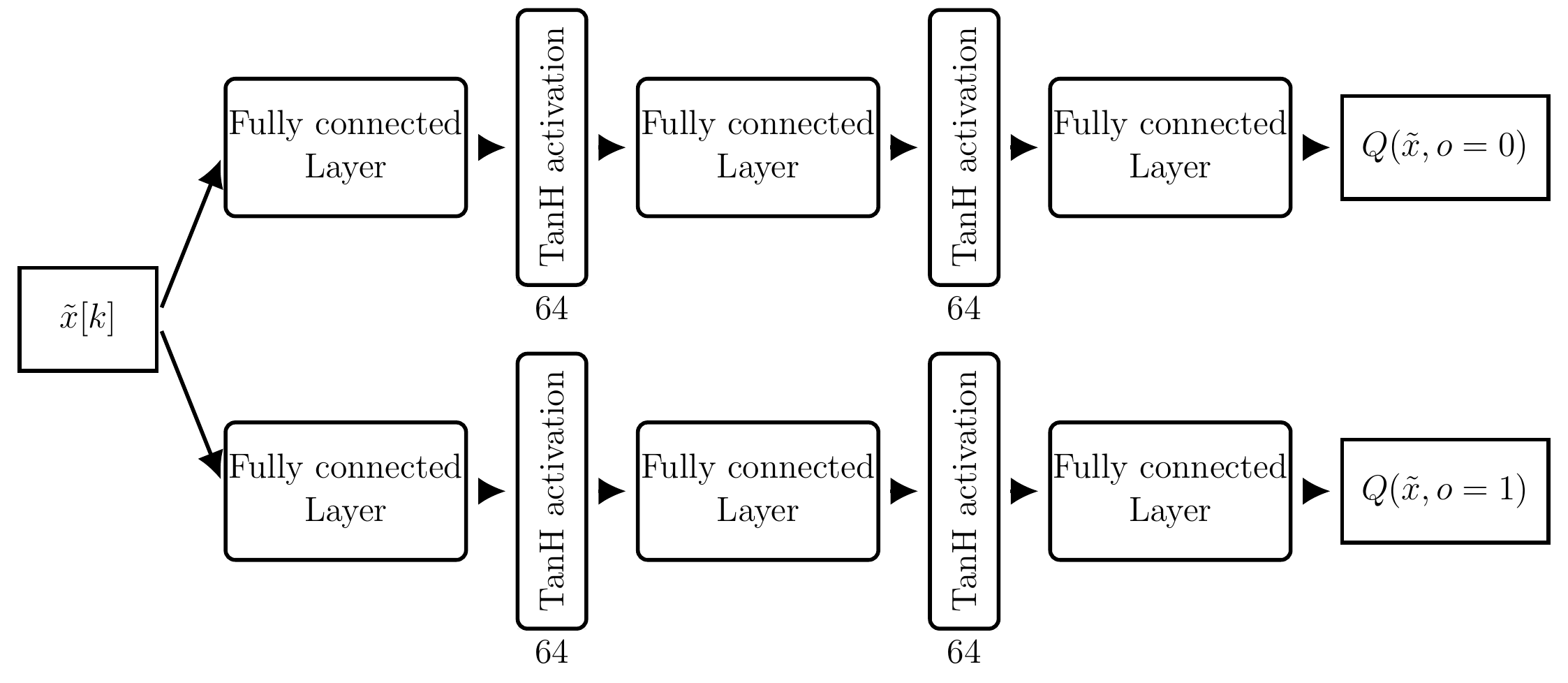}
\caption{Illustration of the parametrization of the Q-function estimator. In this specific implementation, each hidden layer consists of 64 neurons, and the \gls{tanh} activation function is used.}
\label{fig:arch_q_fct}
\end{figure}

As explained in \secref{sec:leveragingrl_etc}, our implementation is based on three main components, the policy over options $\mu(o|\tilde{x})$, the intra option policy $\pi(u|\tilde{x},o)$, and the Q-function $Q(\tilde{x},u)$. \figref{fig:arch_pol_op}, \figref{fig:arch_intra_op}, and \figref{fig:arch_q_fct} show the standard implementations of the respective components for the case of using 2 options. As done in standard \gls{rl}, we normalize the input before it is passed to the networks and clip the output of the intra option policy to reflect the input constraints of the physical system. \figref{fig:arch_intra_op} illustrates that for the case of performing the \gls{zoh}, \ie option 0, the intra option policy does not have to be evaluated. Thus, computational resources can be saved in the forward, as well as the backward pass. For implementing the \gls{nn} estimator of the Q-function (see \figref{fig:arch_q_fct}), we decided to split the estimates for option 0 and option 1 already before the first hidden layer. The reason for this choice is that as the two options are very different, we also expect very different Q-values for the two options, although being in the same state $\tilde{x}$. This is because option 0 is limited to the \gls{zoh}, while option 1 can basically apply any action, depending on the intra option policy. For the same reasons, we arrive at the design choice for the policy over options (see \figref{fig:arch_pol_op}).

For the stability verification and retraining procedure, we apply the following modifications to the previously presented implementation. Instead of using the \gls{tanh} activation function, we apply the \gls{relu} activation. The number of hidden neurons is decreased from 64 to 32. This is due to the fact that at the core of the verification algorithm, we run a modified version of the Simplex algorithm, which simply runs faster if less neurons are used. Further, instead of using the softmax activation function for the policy over options, we calculate $Z=\zeta_0-\zeta_1$. If $Z>0$, this corresponds to performing the \gls{zoh} and choosing option 0, otherwise we use option 1. That way, we achieve deterministic behavior of the policy over options and avoid using the softmax activation function, which is incompatible with the verification framework. Moreover, this deterministic decision is compatible with the stochastic case, as $Z=0$ is equal to $\mu(o=0|\tilde{x})=\mu(o=1|\tilde{x})=\SI{50}{\percent}$, and $Z>0$ corresponds to $\mu(o=0|\tilde{x})>\mu(o=1|\tilde{x})$.

\section{Learning Hierarchical, Periodic Control}
\label{sec:hierarchical_period_ctrl}
Using the proposed algorithm for \gls{etc} is one possibility. However, it is also possible to use it as a normal, hierarchical control algorithm where the two options represent two different \gls{nn} policies that we can sample from. The original PPOC \cite{OptionCriticContPaper} algorithm has also been developed for this case of standard periodic control without any event-triggering involved.

\begin{figure}
\begin{subfigure}{1.0\collen}
	\centering
	\begin{subfigure}[t]{0.54\textwidth}
	\includegraphics[width=\textwidth]{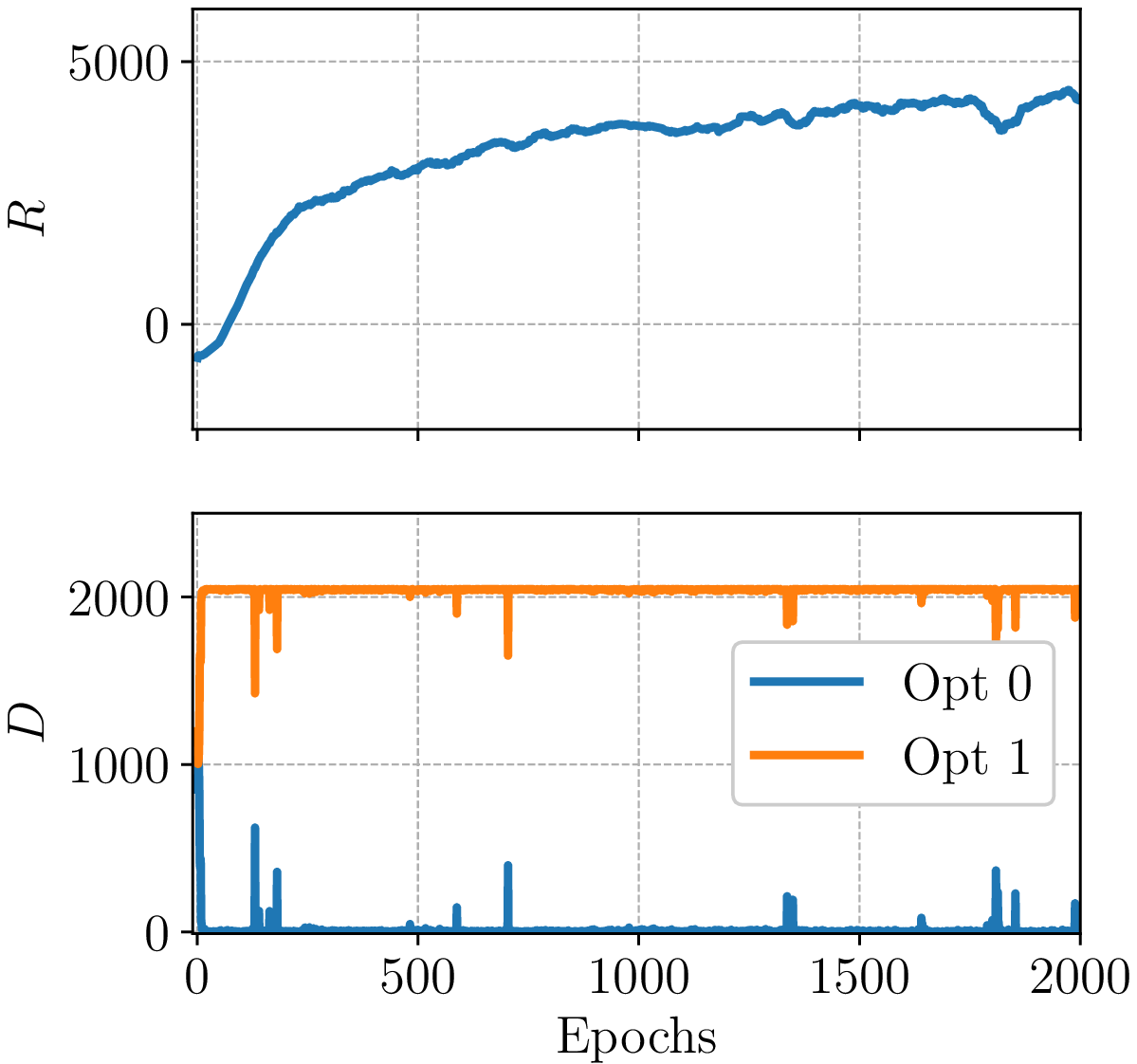}
	\caption{Training progress, using the PPOC algorithm \cite{OptionCriticContPaper}, in the setting of periodic control, without any event-trigger.}
	\label{fig:no_et_ppoc}
	\end{subfigure} \hfill
	\begin{subfigure}[t]{0.36\textwidth}
	\includegraphics[width=\textwidth]{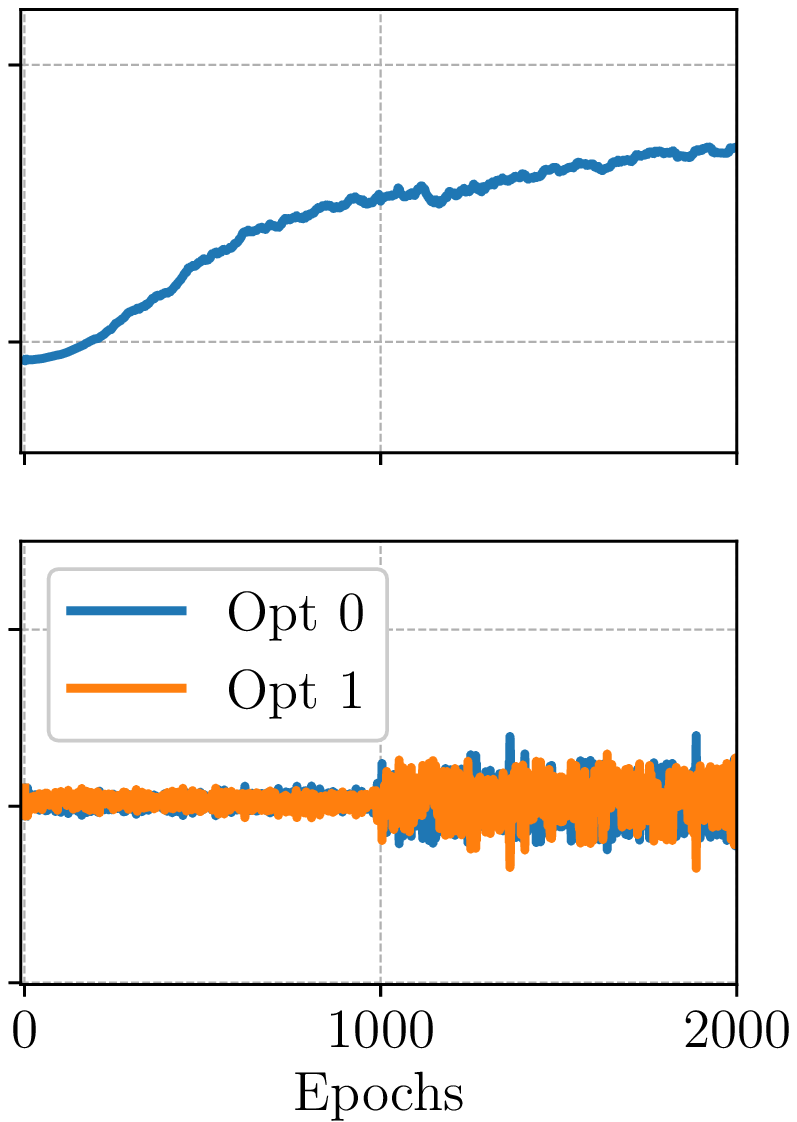}
	\caption{Training progress, using our algorithm, in the setting of periodic control, without any event-trigger.}
	\label{fig:no_et_own}
	\end{subfigure}
\end{subfigure}
\caption{Comparing the training progress for two exemplary policies, one using the PPOC and the other one using our algorithm for the setting of periodic control in the Half-Cheetah environment. Each epoch consists of 2048 ($x,u$)-transitions. The lower plot illustrates how many of those transitions $D$ are conducted using option 0 or option 1.}
\label{fig:no_et_compare_learning}
\end{figure}

Considering this setting, \figref{fig:no_et_compare_learning} exemplarily shows the difference in the learning progress between using the PPOC and the proposed algorithm for the Cheetah environment. As shown in \figref{fig:no_et_ppoc}, the PPOC algorithm quickly collapses to essentially only using one of the options as the other one is almost never executed. In contrast, as illustrated in \figref{fig:no_et_own}, our algorithm ensures that both of the options are used. This is due to the entropy scheduling that prevents the policy over options from being too greedy and also due to the PPO updates. Considering the reward, this might slightly slow down the learning process but is an essential property of the algorithm, which is important when it comes to \gls{etc}. In \gls{etc}, the smooth learning process allows to arrive at policies with intermediate communication savings. On the contrary, we think that the greedy optimization of the \gls{ppoc} algorithm is the reason why its \gls{et} implementation always collapses to one of the extreme cases, either saving no communication at all or never communicating.

\section{Deatils on the Robot Experiments}
\label{sec:supp_material_robot_exp}

For running the robot experiments, three main components are needed: The Apollo robot executing the control actions; the learning agent, running on a computer, using Ubuntu 14.04 together with a Xenomai kernel that sends the control commands to the robot; and the sensors monitoring the robot, which are Apollo's internal sensors and a Vicon camera system. 
Using the Xenomai kernel is essential, as this allows us to check whether timing constraints are violated. 
The Apollo robot is equipped with two KUKA LBR4+ robotic arms. Each of the arms consists of 6 joints. All the experiments presented in this work only use the right arm where at the end, a Barrett Hand is mounted onto the arm. Further details on Apollo can be found in \cite{8263622}. For controlling and simulating the robot, we use the simulation laboratory (SL) framework \cite{SL__2009}. As the SL package is programmed in C, but our learning algorithms are implemented in Python and using Tensorflow, we use a shared memory to exchange information between the Python-based learning pipeline and the C Code, which takes care of the actual robot control.

\end{document}